\documentclass[twocolumn,preprintnumbers,amsmath,amssymb,pra,floatfix]{revtex4}
\usepackage{graphicx}
\usepackage{dcolumn}
\usepackage{bm}
\usepackage{units}%
\usepackage{stmaryrd}%
\usepackage[breaklinks=true,pdfborder={0 0 0}]{hyperref}
\usepackage[usenames]{color}
\usepackage{tikz}
\usepackage{multirow}
\usepackage{mathtools,etoolbox}
\DeclarePairedDelimiterX{\abs}[1]{\lvert}{\rvert}{\ifblank{#1}{{}\cdot{}}{#1}}
\DeclareMathSymbol{\shortminus}{\mathbin}{AMSa}{"39}
\hypersetup{colorlinks=true,citecolor=red}

\definecolor{mblue}{rgb}{0.06, 0.3, 0.57}

\usepackage{physics}
\newcommand\scalemath[2]{\scalebox{#1}{\mbox{\ensuremath{\displaystyle #2}}}}
\begin{document}
\newlength{\LL} \LL 1\linewidth
\title{Quantum-Mechanical Force Balance Between Multipolar Dispersion and \\Pauli Repulsion in Atomic van der Waals Dimers}
\author{Ornella~Vaccarelli}
\email[E-mail: ]{ornella.vaccarelli@uni.lu}
\affiliation{Department of Physics and Materials Science, University of Luxembourg, L-1511 Luxembourg City, Luxembourg}
\author{Dmitry~V.~Fedorov}
\affiliation{Department of Physics and Materials Science, University of Luxembourg, L-1511 Luxembourg City, Luxembourg}
\author{Martin~St\"ohr}
\affiliation{Department of Physics and Materials Science, University of Luxembourg, L-1511 Luxembourg City, Luxembourg}
\author{Alexandre~Tkatchenko}
\email[E-mail: ]{alexandre.tkatchenko@uni.lu}
\affiliation{Department of Physics and Materials Science, University of Luxembourg, L-1511 Luxembourg City, Luxembourg}
\date{\today}

\begin{abstract}
The structure and stability of atomic and molecular systems with van der Waals (vdW) bonding are often determined by the interplay between attractive dispersion interactions and repulsive interactions caused by electron confinement. Arising due to different mechanisms --- electron correlation for dispersion and the Pauli exclusion principle for exchange-repulsion --- these interactions do not appear to have a straightforward connection. In this paper, we use a coarse-grained approach for evaluating the exchange energy for two coupled quantum Drude oscillators and investigate the mutual compensation of the attractive and repulsive forces at the equilibrium distance within the multipole expansion of the Coulomb potential.
This compensation yields a compact formula relating the vdW radius of an atom to its multipole polarizabilities,
$R_{\rm vdW} = A_l^{\,}\, \alpha_l^{\nicefrac{2}{7(l+1)}}$, where $l$ is the multipole rank and $A_l$ is a conversion factor. Such a relation is compelling because it connects an electronic property of an isolated atom (atomic polarizability) with an equilibrium distance in a dimer composed of two closed-shell atoms. We assess the accuracy of the revealed formula for noble-gas, alkaline-earth, and alkali atoms and show that the $A_l$ can be assumed to be universal constants. 
Besides a seamless definition of vdW radii, the proposed relation can also be used for the efficient determination of atomic multipole polarizabilities solely based on the corresponding dipole polarizability and the vdW radius. Finally, our work provides a basis for the construction of efficient and minimally-empirical interatomic potentials by combining multipolar interatomic exchange and dispersion forces on an equal footing.
\end{abstract}

\maketitle

\section{Introduction}
Noncovalent interatomic and intermolecular interactions represent one of the key factors that determine the physicochemical properties of molecules and materials across chemistry, biology and materials science~\cite{Stone2016,Hirschfelder1967,Margenau1971,Kaplan2006}.
Noncovalent interactions are traditionally classified in a perturbative formalism, from which electrostatics, induction, 
Pauli (exchange) repulsion and van der Waals (vdW) dispersion arise as the leading contributions from the first two orders of perturbation theory.
From the perspective of computational modeling, the individual terms are usually treated with different effective approaches. Especially the methods used to describe Pauli repulsion and vdW dispersion typically rely on fundamentally different physical models. The vdW dispersion represents a major part of long-range electron correlation forces arising from Coulomb-coupled instantaneous quantum fluctuations of the electronic charge distribution~\cite{Parsegian2005,Tkatchenko2015,Woods2016,Mahanty1973,Richardson1828,Paranjape1979}.
Common (semi\nobreakdash-)local approximations to density-functional theory (DFT), representing one of the main workhorse methods in atomistic modeling, neglect long-range correlation forces and thus do not account for vdW interactions.
In recent years, an intense effort has been devoted to develop robust approaches to address this challenge~\cite{Klimes2012,Bjorkman2012,Berland2015,Grimme2016,Hermann2017,Stoehr_CSR_2019}.
Although a unified vdW functional valid for all kinds of systems is still under construction~\cite{Hermann2020}, significant progress has been achieved to include dispersion interactions in the form of non-local (vdW) density functionals~\cite{Dobson1999,Dion2004,Vydrov2009,Sabatini2013}.
Furthermore, coarse-grained vdW models have shown great success in describing dispersion interactions at lower computational costs~\cite{Becke2005,Tkatchenko2009,Grimme2010,Odbadrakh2015,Caldeweyher2017}. Among them, the quantum Drude oscillator (QDO) model~\cite{Bade1957,Wang2001,Lamoureux_2003,Lamoureux2003,Sommerfeld2005,Whitfield2006,Jones2013} has been firmly established as an efficient and accurate approach for modeling and understanding vdW interactions~\cite{Whitfield2006,Tkatchenko2012,Jones2013,Reilly2015,Sadhukhan2016,Sadhukhan2017,Hermann_NComm_2017}. Within this approach, each QDO models an atom or a molecule, representing the effective, localized response and polarization fluctuation of its valence electrons.
The success of the coupled-oscillator model is exemplified by its excellent description of the electronic response properties of atoms and molecules. In a continuous formalism, with one oscillator at every point in space, coupled oscillators can describe any response allowed by quantum field theory and thus model the response of arbitrary molecules or materials~\cite{Gobre2016,Ambrosetti2014}.
In the common practical coarse-grained formalism, with each oscillator representing one atom, the QDO framework reproduces the leading-order behavior of the electronic polarizability of atoms~\cite{Tang1968}, providing an accurate and reliable description of polarization effects in molecules and materials~\cite{Thole1981,DiStasio2014}.
Moreover, the QDO model allows one to describe excess electrons in matter~\cite{Wang2001} and to reproduce dispersion-polarized electron densities~\cite{Hermann_NComm_2017} as well as Coulomb interactions between dipolar quantum fluctuations~\cite{Stoehr_NComm_2021}.

Extending the applicability of the QDO framework towards a more complete and systematic description of noncovalent interactions necessitates the incorporation of the exchange-induced repulsion~\cite{VanVleet2016}. Recently, we made a first step in this direction by evaluating the exchange energy between two QDOs within the dipole approximation of the Coulomb potential~\cite{Fedorov2018}.
Here, we take the next natural step by constructing a common coarse-grained approach for the multipolar dispersion and exchange interactions in vdW-bonded atomic or molecular dimers.

It is important to embed our developments of coarse-grained models into the broader context given by the theory of intermolecular interactions for systems composed of nuclei and electrons~\cite{Stone2016}, which states that the equilibrium geometry of two vdW-bonded atoms or molecules
is governed by an interplay of several interactions. The generalized Heitler-London (GHL) theory~\cite{Tang1998} offers one of the most compact schemes for the interatomic energy decomposition. In the GHL approach, isotropic closed-shell atoms only experience mutual exchange-repulsion and dispersion forces. 
Another very successful scheme to describe intermolecular interactions and analyze their complex interplay is based on the symmetry-adapted perturbation theory (SAPT) decomposition~\cite{Jeziorski1994,Hesselmann2005,Szalewicz2012}. The higher-level SAPT methods, while being computationally expensive, approach a ``gold standard'' accuracy~\cite{Parker2014} comparable to the coupled-cluster method with single, double and perturbative triple excitations [CCSD(T)] for small molecules. Within second-order SAPT, which is the most practical approach, one obtains six contributions~\cite{Stone2016,VanVleet2016}:
(i) electrostatics, $E_{\rm elst}^{(1)}$; (ii) exchange, $E_{\rm ex}^{(1)}$; (iii) induction, $E_{\rm ind}^{(2)}$; (iv) exchange-induction, $E_{\rm ex\shortminus ind}^{(2)}$; (v) dispersion, $E_{\rm disp}^{(2)}$; (vi) exchange-dispersion, $E_{\rm ex\shortminus disp}^{(2)}$. Here, the superscripts $(1)$ and $(2)$ denote the order of the perturbation theory required to derive the corresponding term. 
In the case of neutral and isotropic fragments, the two induction contributions, $E_{\rm ind}^{(2)}$ and $E_{\rm ex\shortminus ind}^{(2)}\,$, practically compensate each other~\cite{Hesselmann2014}. Then, the problem reduces to four remaining terms, which still yield significant contributions to
the interaction energy for noble gas dimers~\cite{Shirkov2017}.
On the other hand, the Tang-Toennies (TT) model~\cite{Tang1995}, relying just on the exchange-repulsion and dispersion-attraction interactions, is known to reproduce the binding energy curves of closed-shell dimers with high accuracy and efficiency~\cite{Tang2003}. Recently, an extension (TT2) of this model was proposed~\cite{Sheng2020} to accurately describe noble gas dimers also at relatively short internuclear distances.
Based on the concepts of the GHL theory for interatomic interactions~\cite{Tang1998}, the TT model can be considered as one of the most compact yet accurate models for closed-shell vdW dimers. According to the discussion in Ref.~\cite{Tang1998}, the simplicity of the TT potential arises due to the used analytical asymptotic form of the exchange energy obtained by the surface integral method~\cite{Tang1989,Tang1992}. Since this method is known to deliver the same asymptotic result~\cite{Tang1991} as the approach based on the multipole expansion of the perturbing potential~\cite{Dalgarno1956,Morse1953}, the latter can be used as an alternative way to construct compact TT-like potentials. This idea is supported by our recent study~\cite{Fedorov2018}, which established a quantum-mechanical scaling law, $\alpha_1 \propto R_{\rm vdW}^7$, between the atomic dipole polarizability and the vdW radius from the force balance between exchange-repulsion and dispersion attraction at the equilibrium distance. The corresponding analysis in Ref.~\cite{Fedorov2018} was based on the consideration of these two forces stemming from the dipolar term in the multipole expansion~\cite{Hirschfelder1967,Margenau1971} of the interatomic Coulomb potential. Subsequently, we have derived~\cite{Tkatchenko2020} the proportionality coefficient, which finally led to the relation $\alpha_1 = (4\pi\epsilon_0/a_0^4)\alpha_f^{4/3} R_{\rm vdW}^7$, as expressed in terms of the vacuum permittivity $\epsilon_0$, the Bohr radius $a_0$ and the fine-structure constant $\alpha_f$. Such a relation is not trivial because it connects an electronic polarizability of an atom with an equilibrium distance in a dimer composed of two closed-shell atoms.

In this work, we build on our previous study by going beyond the dipole approximation and considering further terms in the multipole expansion of the interatomic Coulomb potential. This is performed for both exchange and dispersion interactions between closed-shell systems described within the QDO model. To this end, we investigate the balance between the two types of forces, which yield the dominant contributions in vdW-bonded systems. For atomic dimers at the vdW equilibrium distance, this allows us to study a term-by-term compensation of the attractive (dispersion) and repulsive (exchange) forces for each contribution in the multipole expansion of the full Coulomb interaction between the QDOs. This mutual compensation yields a relation between atomic multipole polarizabilities and the vdW radius as first empirically obtained in Ref.~\cite{Fedorov2018}.
The presented relation enables a practical and seamless determination of vdW radii as an effective atomic length scale from atomic polarizabilities.
From the opposite perspective, the generalized relation also allows one to obtain atomic polarizabilities across the periodic table and at arbitrary multipole rank based on (pre-tabulated) vdW radii and without the need to resort to the otherwise challenging direct computation of electronic response properties.
Altogether, our results deliver deeper insights into the connection between Pauli repulsion and dispersion attraction --- two forces which appear at different orders of SAPT. The existence of a quantum-mechanical relation between the two main contributions to the vdW interaction energy at the equilibrium distance reveals a strong connection between exchange and correlation effects and should have implications for achieving an improved understanding of the stability of vdW-bonded matter.

\section{Method: Quantum Drude Oscillators}\label{sec:method}

\begin{figure}[h]
 \includegraphics[width=1.0\linewidth]{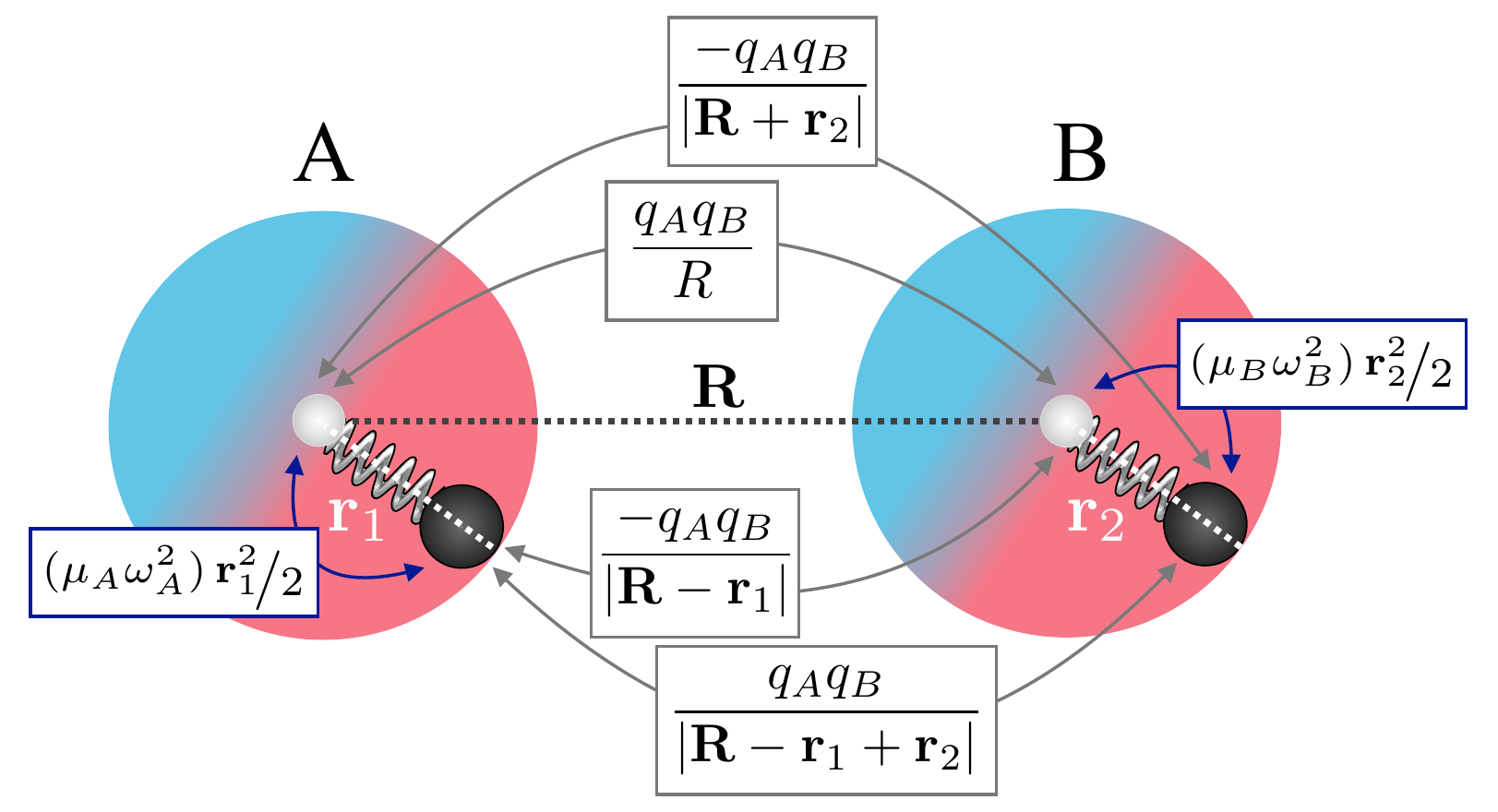}
 \caption{Schematic representation of two QDOs separated by the distance $R = |\mathbf{R}|$. The black and white spheres represent the two Drude particles and fixed nuclei, respectively. The Coulomb interactions between the two QDOs (gray arrows) and the harmonic intra-QDO potentials (blue arrows) are explicitly highlighted with their connection to Eqs.~\eqref{eq.:h_A_and_h_B} and \eqref{eq:V}.}
\label{Fig.:QDOs}
\end{figure}

Let us consider two vdW-bonded atoms, $A$ and $B$, separated by a distance $R$ and describe them within the QDO model, as illustrated in Fig.~\ref{Fig.:QDOs}. Each of the two QDOs representing atoms has three effective parameters --- mass $\mu$, charge $q$ and characteristic frequency $\omega$ --- which are parametrized to reproduce three atomic observables $\lbrace \alpha_1, {\rm C}_6, 
{\rm C}_8\rbrace$~\cite{Jones2013}:
\begin{equation}\label{eq:QDOparam}
  \omega = \frac{4 \, {\rm C}_6 }{3 \, \hbar \, \alpha_1^2} \ ,\ \ \ \mu = \frac{5 \, \hbar \, {\rm C}_6 }{ \omega \, {\rm C}_8 } \ ,\ \ \ q = \sqrt{\mu \omega^2 \alpha_1} \ ,
\end{equation}
where the Drude (quasi-)particle and the related nucleus have charges $(-q)$ and $q$, respectively.
The conditions of Eq.~\eqref{eq:QDOparam} use the dipole polarizability $\alpha_1$ and the dominant dispersion coefficients ${\rm C}_6$ (induced-dipole--induced-dipole interaction) and ${\rm C}_8$ (induced-dipole--induced-quadrupole interactions) in order to parametrize this powerful model, able to efficiently reproduce long-range forces and electronic response properties of atoms and molecules. 

The Hamiltonian of the interacting QDOs is given by $H = H_0 + V$, where $V$ is the interaction and $H_0 = h_A + h_B$
consists of the unperturbed QDO Hamiltonians
\begin{equation}\label{eq.:h_A_and_h_B}
\begin{split}
h_A(\textbf{r}) &= -({\hbar^2}/{2\mu_A}) \boldsymbol{\nabla}_{\textbf{r}}^2 + ({\mu_A \omega_A^2}/{2}) \textbf{r}^2\ , \\
h_B(\textbf{r}) &= -({\hbar^2}/{2\mu_B}) \boldsymbol{\nabla}_{\textbf{r}}^2 + ({\mu_B \omega_B^2}/{2}) (\textbf{r} - \textbf{R})^2\ .
\end{split}
\end{equation}
The corresponding wave functions are given by
\begin{equation}
\begin{split}
 \psi_A(\textbf{r}) & = ({\mu_A\omega_A}/{\pi\hbar})^{\nicefrac{3}{4}}\,
 e^{- \frac{\mu_A \omega_A}{2 \hbar} \textbf{r}^2}\ , \\
 \psi_B(\textbf{r}) &= ({\mu_B\omega_B}/{\pi\hbar})^{\nicefrac{3}{4}}\,
 e^{- \frac{\mu_B \omega_B}{2 \hbar}
 (\textbf{r} - \textbf{R})^2}\ .
\end{split}
\end{equation}
The full Coulomb interaction between the two QDOs is
\begin{equation}\label{eq:V}
V = \tfrac{q_A q_B}{4\pi\epsilon_0} \left\{ \tfrac{1}{R} + \tfrac{1}{\abs{\textbf{R} - \textbf{r}_1 + \textbf{r}_2 } } - \tfrac{1}{\abs{\textbf{R} - \textbf{r}_1}} - \tfrac{1}{\abs{\textbf{R} + \textbf{r}_2}} \right\}\ ,
\end{equation}
where $\textbf{r}_1$ and $\textbf{r}_2$ are the coordinates of the Drude particles measured from the corresponding fixed nuclei. 
The Coulomb interaction can be written as a multipole expansion as~\cite{Margenau1938}
\begin{equation}
V = \textstyle\sum\limits_{n=1,2,...} V_n = V_1 + V_2 + V_3 + V_4 + V_5 + ...\ ,
\label{eq:expansion_5}
\end{equation}
where $V_n \propto R^{-(n+2)}$ with $R = \abs{\textbf{R}}$. Furthermore, $n = l_A + l_B - 1$, where $l_A$ and $l_B$ refer to the rank of the multipole moments of the two interacting QDOs. Here, we restrict our consideration to the first five terms in the multipole expansion of Eq.~\eqref{eq:expansion_5}.
The first term, $V_1 \propto R^{-3}$, corresponds to the dipole approximation of the Coulomb potential, 
$V_1 \equiv V_{{\rm dip}} = \frac{q_A q_B}{4\pi\epsilon_0} \left( \tfrac{(\textbf{r}_1 {{\cdot}} \textbf{r}_2)}{R^3} \!-\! \tfrac{3(\textbf{r}_1 {{\cdot}} \textbf{R}) (\textbf{r}_2 {{\cdot}} \textbf{R}) }{R^5} \right)$,
 describing the dipole-dipole (d-d) electrostatic interaction ($l_A=l_B=1$). The higher terms arise from the dipole-quadrupole (d-q) for $V_2 \propto R^{-4}$, dipole-octupole (d-o) + quadrupole-quadrupole (q-q) for $V_3 \propto R^{-5}$, dipole-hexadecapole (d-h) + quadrupole-octupole (q-o) for $V_4 \propto R^{-6}$ and dipole-triakontadipole (d-t) + quadrupole-hexadecapole (q-h) + octupole-octupole (o-o) interactions for $V_5 \propto R^{-7}\,$. The formulas for $V_n\,$, with $n$ = 2, 3, 4 and 5, are given in Appendix~\ref{appendix:A}. Within the next section, we consider the multipolar contributions to the dispersion and exchange interaction between two QDOs. 
 The analytical formulas are derived in the most general form valid in any system of units, whereas we employ (a.u.) atomic units (with $4\pi\epsilon_0 = \hbar = 1$), to present our numerical results in Section \ref{sec:num_results}.

\section{Results}

\subsection{Dispersion interaction}

The multipole expansion has been the starting point for quantum-mechanical perturbation calculations of the vdW dispersion interactions of Coulomb-coupled Drude oscillators~\cite{Wang2001,Hermann2017,Crosse2009}. Owing to this approach, the vdW dispersion energies can be expressed in terms of the atomic multipole polarizabilities (with $l=1, 2, ...$) 
\begin{equation}\label{eq:alpha_QDO}
\alpha_l \equiv
\alpha_{l,{\rm QDO}} = \left( \tfrac{q^2}{\mu \omega^2} \right) \tfrac{(2 l - 1) !! }{l} \left( \tfrac{\hbar}{2 \mu \omega } \right)^{l-1}
\end{equation}
by using the series expansion~\cite{Jones2013}
\begin{equation}
E^{{\rm AB,disp}} \!=\! - \sum_{l_A l_B} \abs{T_{l_A,\, l_B}^{AB}}^2\, \tfrac{\alpha_{l_A}^A \alpha_{l_B}^B}{(4\pi\epsilon_0)^2} \left[ \tfrac{\hbar}{4} \tfrac{l_A l_B \omega_A \omega_B}{ ( l_A \omega_A + l_B \omega_B) } \right]\ ,
\end{equation}
where $T_{l_A, \, l_B}^{AB}$ represents the multipole--multipole interaction-tensor.
We remark that $T_{l_A, \, l_B}^{AB}$ above has been obtained using a spherical harmonic expansion of the Coulomb potential instead of the Cartesian multipolar potential described in Appendix A. Both expansions yield equivalent results~\cite{Smith1998,Lin2015}.
In the Supplemental Material of Ref.~\cite{Jones2013}, the following spherical components of this tensor were given
\begin{equation}
\begin{split}
|T_{1,1}|^2 &= 6 R^{-6} \, \quad ,~~~\, |T_{1,2}|^2 = 15 R^{-8} ~~\, ,\\ |T_{1,3}|^2 &= 28 R^{-10}\ ,~~~\,|T_{2,2}|^2 = 70 R^{-10}\ .
\end{split}
\label{eq:T_ll_Jones}
\end{equation}
For our derivations here, we further introduce the higher-order coupling components via a generalized expression of the multipolar interaction tensor,
\begin{equation}
\abs{T_{l_A,\, l_B}^{AB}}^2 \equiv \sum\limits_{m_A = -l_A}^{l_A}\, \sum\limits_{m_B = - l_B}^{l_B} |T_{l_A m_A,\, l_B m_B} (\mathbf{R})|^2\ .
\end{equation}
This tensor is derived based on the approach of Ref.~\cite{Popelier2001} used by some of us in Ref.~\cite{Kleshchonok2018} as well. 
Popelier~\emph{et al.}~\cite{Popelier2001} employed the relation
\begin{equation}
\begin{split}
T&_{l_A m_A,\, l_B m_B} (\mathbf{R}) = (-1)^{l_A}\, \sqrt{\tfrac{(2 l_A + 2 l_B + 1)!}{(2 l_A)! (2 l_B)!}} \\
& \times \scalemath{0.8}{
\setlength\arraycolsep{1.5pt}
\begin{pmatrix} l_A & l_B & l_A + l_B
\\ m_A & m_B & -(m_A + m_B) \end{pmatrix} }\,
I_{l_A + l_B\, ,\, -(m_A + m_B)} (\mathbf{R})\ ,
\end{split}
\label{eq:T_Popelier}
\end{equation}
where the expression in the large parentheses is a Wigner $3j$-symbol~\cite{Varshalovich1988} and the irregular normalized spherical harmonics are
\begin{equation}
I_{l, m} (\mathbf{r}) = \sqrt{\tfrac{4\pi}{2 l +1}}\, r^{-l-1}\, Y_{l, m} (\theta, \phi) \, .
\end{equation}
The spherical harmonics are defined as~\cite{Varshalovich1988}
\begin{equation}
Y_{l, m} (\theta, \phi) = \sqrt{\tfrac{2 l +1}{4\pi}}\sqrt{\tfrac{(l-m)!}{(l+m)!}}\, P_l^m (\cos \theta)\, e^{i m \phi}\ ,
\end{equation}
where $P_l^m (\cos \theta)$ are the associated Legendre polynomials. If we assume now that the distance between
two atoms is along the $z$ axis, $\mathbf{R} = (0, 0, R)$, then $\cos \theta = 1$. Due to $P_l^m (1) = \delta_{m, 0}\,$, one can then easily obtain
\begin{equation}
Y_{l, m}^* (0, \phi)\, Y_{l, m} (0, \phi) = \tfrac{2 l +1}{4\pi} \tfrac{(l-m)!}{(l+m)!} \delta_{m, 0}\ .
\end{equation}
Consequently, we have
\begin{equation}
 \abs{T_{l_A,\, l_B}^{AB}}^2 = ~\tfrac{(2 l_A + 2 l_B + 1)! \sum\limits_{m = -l_{\le}}^{m = +l_{\le}} \scalemath{0.8}{
\setlength\arraycolsep{1.5pt}
\begin{pmatrix} l_A & l_B & l_A + l_B
\\ m & -m & 0 \end{pmatrix}^2 }}{(2 l_A)! \, (2 l_B)! \, R^{2 (l_A + l_B + 1)} } \ , 
\end{equation}
where $l_{\le} = \{ l_A\, ,\ {\rm if}\ l_A \le l_B\, ;\ 
l_B\, ,\ {\rm if}\ l_B \le l_A \}$.
Now we use the following property of the Wigner $3j$-symbols~\cite{Varshalovich1988}
\begin{equation}
\begin{split}
& \scalemath{0.9}{
\setlength\arraycolsep{1.5pt}
\begin{pmatrix} j_1 & j_2 & j_1 + j_2
\\ m_1 & m_2 & -(m_1 + m_2) \end{pmatrix} } = (-1)^{j_1 - j_2 + m_1 + m_2} \\ & \quad
\sqrt{\tfrac{(2 j_1)! (2 j_2)! (j_1 + j_2 + m_1 + m_2)! (j_1 + j_2 - m_1 - m_2)!}
{(2 j_1 + 2 j_2 + 1)! (j_1 + m_1)! (j_1 - m_1)! (j_2 + m_2)! (j_2 - m_2)!}}\ \ ,
\end{split}
\end{equation}
which gives us
\begin{equation}
\abs{T_{l_A,\, l_B}^{AB}}^2 = \tfrac{ \sum\limits_{m = - l_{\le}}^{m = + l_{\le}}
\frac{(l_A + l_B)! (l_A + l_B)!}{(l_A + m)! (l_A - m)! (l_B + m)! (l_B - m)!}\ }{R^{2 (l_A + l_B + 1)}}\ .
\end{equation}
Obviously, $|T_{l_A,\, l_B}^{AB}|^2 = |T_{l_B,\ l_A}^{AB}|^2$. Therefore, it is enough to derive the components with
$l_A \le l_B$. This means
\begin{align}
\abs{T_{l_A,\, l_B}^{AB}}^2 = \frac{ \sum\limits_{m = 0}^{m = + l_A}
(2 - \delta_{m, 0}) \scalemath{0.8}{
\setlength\arraycolsep{1.5pt}
\begin{pmatrix} l_A + l_B \\ l_A - m \end{pmatrix} }
\scalemath{0.8}{
\setlength\arraycolsep{1.5pt}
\begin{pmatrix} l_A + l_B \\ l_B - m \end{pmatrix} } \ }{ R^{2 (l_A + l_B + 1)} }\ ,
\end{align}
where the factorials have been rewritten in terms of the binomial coefficients
$\scalemath{0.62}{\begin{pmatrix} n \\ k \end{pmatrix} } = \frac{n!}{(n-k)!\, k!}$\ . Then, we obtain
\begin{align}
\begin{split}
 |T_{1,4}|^2 &= 45 R^{-12} \, \quad ,\ \ \ |T_{1,5}|^2 = 66 R^{-14} \quad ,\\
 |T_{2,3}|^2 &= 210 R^{-12} ~\, \, , \ \ \ |T_{2,4}|^2 = 495 R^{-14} ~\, ,\\
 |T_{3,3}|^2 &= 924 R^{-14}\ \ ,
\end{split}
\end{align}
in addition to the results of Eq.~\eqref{eq:T_ll_Jones}.
With the above expressions, the first few multipolar contributions to the dispersion energy between two oscillators become
\begin{align}
\label{eq:disp_en}
E_{1({\rm d \shortminus d})}^{{\rm AB,disp}} &= \nonumber -\tfrac{3 k_e^2}{2 R^6}\, \alpha_1^A \alpha_1^B \tfrac{\hbar \omega_A \omega_B }{ \omega_A + \omega_B }\ , \\
E_{2({\rm d \shortminus q})}^{{\rm AB,disp}} &= \nonumber -\tfrac{15 k_e^2}{2 R^8} \left[ \alpha_1^A \alpha_2^B \tfrac{\hbar \omega_A \omega_B }{ \omega_A + 2 \omega_B } + \alpha_2^A \alpha_1^B \tfrac{\hbar \omega_A \omega_B }{ 2 \omega_A + \omega_B } \right]\ , \\
E_{3({\rm d \shortminus o})}^{{\rm AB,disp}} &= \nonumber -\tfrac{21 k_e^2}{ R^{10} } \left[ \alpha_1^A \alpha_3^B \tfrac{\hbar \omega_A \omega_B }{ \omega_A + 3 \omega_B } + \alpha_3^A \alpha_1^B \tfrac{\hbar \omega_A \omega_B }{ 3 \omega_A + \omega_B } \right]\ , \\
E_{3({\rm q \shortminus q})}^{{\rm AB,disp}} &= \nonumber -\tfrac{35 k_e^2}{ R^{10}}\, \alpha_2^A \alpha_2^B \tfrac{\hbar \omega_A \omega_B }{ \omega_A + \omega_B } \ , \\
E_{4({\rm d \shortminus h})}^{{\rm AB,disp}} &= \nonumber -\tfrac{45 k_e^2}{ R^{12} } \left[ \alpha_1^A \alpha_4^B \tfrac{\hbar \omega_A \omega_B }{ \omega_A + 4 \omega_B } + \alpha_4^A \alpha_1^B \tfrac{\hbar \omega_A \omega_B }{ 4 \omega_A + \omega_B } \right] \ , \\
E_{4({\rm q \shortminus o})}^{{\rm AB,disp}} &= \nonumber -\tfrac{315 k_e^2}{ R^{12} } \left[ \alpha_2^A \alpha_3^B \tfrac{\hbar \omega_A \omega_B }{ 2 \omega_A + 3 \omega_B } + \alpha_3^A \alpha_2^B \tfrac{\hbar \omega_A \omega_B }{ 3 \omega_A + 2 \omega_B } \right] \ , \\
E_{5({\rm d \shortminus t})}^{{\rm AB,disp}} &= \nonumber -\tfrac{165 k_e^2}{2 R^{14} } \left[ \alpha_1^A \alpha_5^B \tfrac{\hbar \omega_A \omega_B }{ \omega_A + 5 \omega_B } + \alpha_5^A \alpha_1^B \tfrac{\hbar \omega_A \omega_B }{ 5 \omega_A + \omega_B } \right] \ , \\
E_{5({\rm q \shortminus h})}^{{\rm AB,disp}} &= \nonumber -\tfrac{495 k_e^2}{ R^{14} } \left[ \alpha_2^A \alpha_4^B \tfrac{\hbar \omega_A \omega_B }{ \omega_A + 2 \omega_B } + \alpha_4^A \alpha_2^B \tfrac{\hbar \omega_A \omega_B }{ 2 \omega_A + \omega_B} \right] \ , \\
E_{5({\rm o \shortminus o})}^{{\rm AB,disp}} &= -\tfrac{693 k_e^2}{R^{14}}\, \alpha_3^A \alpha_3^B \tfrac{\hbar \omega_A \omega_B }{ \omega_A + \omega_B }\ \ ,
\end{align}
where $k_e = (4\pi\epsilon_0)^{-1}$ is the Coulomb constant.

Based on the above formulas, we can now rewrite the dispersion energy in its conventional expansion~\cite{Stone2016,Jones2013}
\begin{equation}
\label{eq:disp_en_conv}
  E^{{\rm AB,disp}} = \!\!\! \textstyle\sum\limits_{n=1,2,...} \!\!\! E_n^{{\rm AB,disp}} = \!\!\! \textstyle\sum\limits_{n = 1,2,...} \!\!\! - \tfrac{{\rm C}_{(2n+4)}^{{\rm AB}}}{R^{(2n+4)}}\ \ ,
\end{equation}
where ${\rm C}_{(2n+4)}^{{\rm AB}}$ are the dispersion coefficients and all the contributions to $E_n^{{\rm AB,disp}}$, with $n$ up to 5, are given by Eq.~\eqref{eq:disp_en}. 
Equation~\eqref{eq:disp_en_conv} arises from second-order perturbation theory with the multipole expansion of the Coulomb potential, as an interaction potential between spherically symmetric atoms. 
The leading term is the dipole-dipole (d-d) interaction, $E_1^{{\rm AB,disp}}\propto R^{-6}$, stemming from the dipolar potential, $V_{{\rm dip}} \propto R^{-3}$. 
The higher-order terms in the multipole expansion of the Coulomb interaction yield the dispersion energies $E_2^{{\rm AB,disp}} \propto R^{-8}$, $E_3^{{\rm AB,disp}} \propto R^{-10}$, $E_4^{{\rm AB,disp}} \propto R^{-12}$ and $E_5^{{\rm AB,disp}} \propto R^{-14}$ coming, respectively, from the instantaneous dipole-quadrupole (d-q), dipole-octupole (d-o) and quadrupole-quadrupole (q-q), dipole-hexadecapole (d-h) and quadrupole-octupole (q-o) interactions, and dipole-triakontadipole (d-t), quadrupole-hexadecapole (q-h) and octupole-octupole (o-o) interactions.  
For non-centrosymmetric molecules, Eq.~\eqref{eq:disp_en_conv} would have terms
with odd powers in $R$ starting with $\propto R^{-7}$~\cite{Stone2016}.
However, here we restrict our consideration to vdW-bonded atoms assumed to possess closed valence-electron shells with a spherically-symmetric charge density, for which the dispersion terms proportional to $1/R^{2i+1}$, with $i \in \mathbb{N}$, vanish.

\subsection{Exchange-repulsion interaction}

The above derivation of the dispersion energy was performed for the general case of two QDOs with arbitrary parameters, however the description of the exchange-repulsion between two QDOs is more subtle. 
The exchange interaction should obviously be present for two different QDOs, as caused by the Pauli repulsion between electrons constituting the two Drude particles. Nonetheless, in order to construct the exchange interaction, one needs to deal with indistinguishable particles, a concept that requires generalization for two Drude particles possessing different parameters. Our starting assumption is that the exchange energy should be proportional to the overlap integral $S$ between the wave functions of two different QDOs, similar to the case of two identical Drude particles~\cite{Fedorov2018}. This assumption was recently employed~\cite{Silvestrelli2019} for a simplified generalization of the coarse-grained dipole-dipole exchange energy of a homonuclear dimer, $E^{{\rm ex}}_{({\rm d \shortminus d})} \approx k_e q^2 S / 2 R$, derived in Ref.~\cite{Fedorov2018}.
The authors of Ref.~\cite{Silvestrelli2019} have simply replaced the overlap integral $S$ of two identical QDOs by its counterpart obtained for different QDOs and shown that already such a simplified treatment improves their computational scheme for vdW dispersion interactions. However, due to the coarse-grained treatment of valence electrons within the QDO model, care needs to be taken for the most general definition of the exchange energy between QDOs. This is a subject of our ongoing studies. Here, we follow the approach of Ref.~\cite{Fedorov2018} and derive multipole contributions to the exchange energy of two identical QDOs.

Formally, we consider two indistinguishable Drude particles ($\mu =\mu_A =\mu_B\,$, $\omega =\omega_A =\omega_B\,$ and $q=q_A= q_B$) as bosons assuming that they represent closed valence shells with vanishing total spin. Therefore, the total wave function of a dimer should be written as a permanent
\begin{equation}
\Psi(\textbf{r}_1, \textbf{r}_2) = \tfrac{1}{\sqrt{2}} \big( \psi_A(\textbf{r}_1)\psi_B(\textbf{r}_2) + \psi_A(\textbf{r}_2)\psi_B(\textbf{r}_1 ) \big)\ .
\end{equation}
By employing the Heitler-London perturbation theory~\cite{Heitler1927,Slater1965}, the exchange energy for two identical vdW-bonded QDOs at their equilibrium distance becomes well approximated with its exact asymptotic result
given by the exchange integral~\cite{Fedorov2018}
\begin{align}
\label{eq:ex_integral}
J^{{\rm ex}} = \mel{\psi_A(\textbf{r}_1) \psi_B(\textbf{r}_2) }{V}{ \psi_A(\textbf{r}_2) \psi_B(\textbf{r}_1)}\ .
\end{align}
The evaluation of Eq.~\eqref{eq:ex_integral} with the expansion of Eq.~\eqref{eq:expansion_5} results in multipole contributions to the exchange energy, where each of them is directly proportional to the overlap integral defined as
\begin{equation}
S = \abs{ \braket{ \psi_A }{ \psi_B } }^2 = e^{-\frac{\mu \omega}{2 \hbar} R^2}\ .
\end{equation}
For the dipole-dipole contribution, $V_1 \propto R^{-3}$, we obtain
\begin{equation}
\label{eq:J_ex_d_d}
J_{1({\rm d \shortminus d})}^{{\rm ex}} = \tfrac{k_e q^2 S}{2 R}\ ,
\end{equation}
which reproduces the result of Ref.~\cite{Fedorov2018}.
Now, we evaluate further contributions going beyond the dipole approximation. For the dipole-quadrupole interaction, described by the second term, $V_2 \propto R^{-4}$, in the multipole expansion of the Coulomb potential, we derive 
\begin{equation}
\label{eq:J_ex_d_q}
J^{{\rm ex}}_{2({\rm d \shortminus q})} = \tfrac{3 k_e q^2 S }{ 4 R }\ .
\end{equation}
Then, the next term, $V_3 \propto R^{-5}$, has two contributions, 
$V_{3 ({\rm d \shortminus o})}$ and $V_{3 ({\rm q \shortminus q})}\,$, related to
the dipole-octupole and the quadrupole-quadrupole interaction~\cite{Margenau1971,Merwe1967}, respectively.
The corresponding exchange integrals are obtained as
\begin{equation}
\begin{split}
J^{{\rm ex}}_{3({\rm d \shortminus o})} & = \tfrac{k_e q^2 S }{ 2 R }\ , \\
J^{{\rm ex}}_{3({\rm q \shortminus q})} & = \tfrac{ 3 k_e q^2 S }{ 8 R } \left( 1 \!-\! \tfrac{\hbar}{R^2 \mu \omega} \!-\! \tfrac{\hbar^2}{R^4 \mu^2 \omega^2}\right) .
\end{split}
\end{equation}
Further on, we have two contributions from $V_4 \propto R^{-6}$, the dipole-hexadecapole (d-h) and the quadrupole-octupole (q-o) interaction. The related exchange integrals are
\begin{equation}
\begin{split}
J^{{\rm ex}}_{4({\rm d \shortminus h})} & = \tfrac{ 5 k_e q^2 S }{ 16 R } \left(1 \!-\! \tfrac{3 \hbar}{R^2 \mu \omega } \!-\! \tfrac{9 \hbar^2}{R^4 \mu^2 \omega^2 } \right)\ ,\\
J^{{\rm ex}}_{4({\rm q \shortminus o})} &=
\tfrac{ 5 k_e q^2 S }{ 8 R }\ .
\end{split}
\end{equation}
Finally, for the dipole-triakontadipole (d-t), quadrupole-hexadecapole (q-h) and octupole-octupole (o-o) interactions, from $V_5 \propto R^{-7}$, we obtain
\begin{align}
J^{{\rm ex}}_{5({\rm d \shortminus t})} &= \nonumber \tfrac{ 3 k_e q^2 S }{ 16 R } \left(1 \!-\! \tfrac{15 \hbar}{R^2 \mu \omega } \!-\! \tfrac{105 \hbar^2}{R^4 \mu^2 \omega^2 } \right) \,\\ 
J^{{\rm ex}}_{5({\rm o \shortminus o})} &= \tfrac{ 5 k_e q^2 S }{ 16 R }\, , \\
J^{{\rm ex}}_{5({\rm q \shortminus h})} &= \nonumber \tfrac{ 15 k_e q^2 S }{ 32 R } \left(1 \!-\! \tfrac{7 \hbar}{4 R^2 \mu \omega } \!-\! \tfrac{35 \hbar^2}{ 4 R^4 \mu^2 \omega^2 } \!-\! \tfrac{21 \hbar^3}{2 R^6 \mu^3 \omega^3 } \right) .
\end{align}
According to Eqs.~(\ref{eq:J_ex_d_d}) and (\ref{eq:J_ex_d_q}), $J^{{\rm ex}}_{2({\rm d \shortminus q})}$ is larger than $J^{{\rm ex}}_{1({\rm d \shortminus d})}$ for all interatomic distances. This is in contrast to the dispersion contributions, where $E^{{\rm disp}}_{1({\rm d \shortminus d})}$ clearly dominates at large distances. Such a nonmonotonic behavior of the multipole contributions, as we obtain here for the exchange energy, was also found in Ref.~\cite{Chalasinski1977} for the multipole expansion of the exchange-dispersion energy. 

Now, we will use the derived dominant multipole contributions to the dispersion and exchange energies, in order to study the balance of the corresponding forces at the equilibrium distance in homonuclear dimers.

\subsection{Force balance between multipolar dispersion and exchange contributions}

The equilibrium geometry of atomic or molecular systems is dictated by the condition that the net forces acting on each atom vanish. Therefore, for two atoms or molecules separated by a distance $R$, this condition is determined by
$\mathbf{F}_{\rm net} (R_{\rm eq}) = -\boldsymbol\nabla_\mathbf{R} E_{\rm tot} (R)|_{R=R_{\rm eq}} = 0$, where $R_{\rm eq}$ represents the equilibrium distance and $E_{\rm tot}$ is the total interaction energy. 
The structure, stability and dynamics of vdW-bonded atomic dimers are governed by the interplay between the dispersion and exchange interactions~\cite{Tang1998}. This means that at $R=R_{\rm eq}\,$ the two respective forces have to mutually compensate each other. In what follows, we consider such a compensation by going beyond the dipole approximation for the interaction, in order to obtain higher-order multipole contributions to the attractive and repulsive forces.

At the equilibrium distance, $R_{{\rm eq}} = 2 R_{\rm vdW}$, in homonuclear dimers composed of two identical Drude particles ($\mu = \mu_A = \mu_B$, $\omega = \omega_A = \omega_B$ and $q = q_A = q_B$), the exchange force can be well approximated by the expression $F^{{\rm ex}} \approx - \nabla_R\, J^{{\rm ex}}$, according to Ref.~\cite{Fedorov2018} and our discussion above. In addition, at the internuclear distances comparable to or larger than the equilibrium one,
$R \gg \sqrt{\hbar/\mu\omega}$~\cite{Fedorov2018}.
Then, the corresponding multipole contributions to the exchange force are obtained as
\begin{alignat}{4}
\label{eq:exchange_forces}
F_{1({\rm d \shortminus d})}^{{\rm ex}} &\nonumber\approx \tfrac{ \alpha_1 \hbar\omega S}{2 (4\pi\epsilon_0)} \left(\tfrac{\mu \omega}{\hbar} \right)^2\ &&, \ \ F_{2({\rm d \shortminus q})}^{{\rm ex}} &&\approx \tfrac{ 3 \alpha_1 \hbar \omega S}{4 (4\pi\epsilon_0)} \left( \tfrac{\mu \omega}{\hbar} \right)^2\ &&, \\
F_{3({\rm d \shortminus o})}^{{\rm ex}} &\nonumber\approx \tfrac{ \alpha_1 \hbar\omega S}{2 (4\pi\epsilon_0)} \left(\tfrac{\mu \omega}{\hbar} \right)^2\ &&, \ \ F_{3({\rm q \shortminus q})}^{{\rm ex}} &&\approx \tfrac{ 3 \alpha_1 \hbar \omega S}{8 (4\pi\epsilon_0)} \left( \tfrac{\mu \omega}{\hbar} \right)^2\ &&, \\
F_{4({\rm d \shortminus h})}^{{\rm ex}} &\nonumber\approx \tfrac{5 \alpha_1 \hbar\omega S}{16 (4\pi\epsilon_0)} \left(\tfrac{\mu\omega}{\hbar} \right)^2\ &&, \ \ F_{4({\rm q \shortminus o})}^{{\rm ex}} &&\approx \tfrac{ 5 \alpha_1 \hbar \omega S}{8 (4\pi\epsilon_0)} \left( \tfrac{\mu \omega}{\hbar} \right)^2\ &&, \\
F_{5({\rm d \shortminus t})}^{{\rm ex}} &\nonumber\approx \tfrac{3 \alpha_1 \hbar\omega S}{16 (4\pi\epsilon_0)} \left(\tfrac{\mu\omega}{\hbar} \right)^2\ &&, \ \ F_{5({\rm q \shortminus h})}^{{\rm ex}} &&\approx \tfrac{ 15 \alpha_1 \hbar \omega S}{32 (4\pi\epsilon_0)} \left( \tfrac{\mu \omega}{\hbar} \right)^2\ &&, \\
F_{5({\rm o \shortminus o})}^{{\rm ex}} & \approx \tfrac{5 \alpha_1 \hbar \omega S}{16 (4\pi\epsilon_0)} \left( \tfrac{\mu \omega}{\hbar} \right)^2\ &&.
\end{alignat}
From Eq.~\eqref{eq:disp_en} we calculate the multipole contributions to the dispersion force (for homonuclear dimers)
\begin{alignat}{4}
\label{eq:dispersion_forces}
F_{1({\rm d \shortminus d})}^{{\rm disp}} &=\nonumber -\tfrac{9 \alpha_1 \alpha_1 \hbar \omega}{2 R^7 (4\pi\epsilon_0)^2} \ &&,\ \ 
F_{2({\rm d \shortminus q})}^{{\rm disp}} &&= -\tfrac{40 \alpha_1 \alpha_2 \hbar \omega}{R^9 (4\pi\epsilon_0)^2}\ &&, \\
F_{3({\rm d \shortminus o})}^{{\rm disp}} &=\nonumber -\tfrac{105 \alpha_1 \alpha_3 \hbar \omega}{R^{11} (4\pi\epsilon_0)^2}\ &&,\ \ 
F_{3({\rm q \shortminus q})}^{{\rm disp}} &&= -\tfrac{175 \alpha_2 \alpha_2 \hbar \omega}{R^{11} (4\pi\epsilon_0)^2}\ &&, \\
F_{4({\rm d \shortminus h})}^{{\rm disp}} &=\nonumber -\tfrac{216 \alpha_1 \alpha_4 \hbar \omega}{R^{13} (4\pi\epsilon_0)^2}\ &&,\ \ 
F_{4({\rm q \shortminus o})}^{{\rm disp}} &&= -\tfrac{1512 \alpha_2 \alpha_3 \hbar \omega}{R^{13} (4\pi\epsilon_0)^2}\ &&, \\
F_{5({\rm d \shortminus t})}^{{\rm disp}} &=\nonumber -\tfrac{385 \alpha_1 \alpha_5 \hbar \omega}{R^{15} (4\pi\epsilon_0)^2}\ &&,\ \ 
F_{5({\rm q \shortminus h})}^{{\rm disp}} &&= -\tfrac{4620 \alpha_2 \alpha_4 \hbar \omega}{R^{15} (4\pi\epsilon_0)^2}\ &&, \\
F_{5({\rm o \shortminus o})}^{{\rm disp}} &= - \tfrac{4851 \alpha_3 \alpha_3 \hbar \omega}{R^{15} (4\pi\epsilon_0)^2}\ &&.
\end{alignat}
At $R = R_{\rm eq}$, the attractive and repulsive forces should cancel each other. Within the dipole approximation, from the force balance, $F_{1({\rm d \shortminus d})}^{{\rm disp}} + F_{1({\rm d \shortminus d})}^{{\rm ex}} = 0$, one obtains
\begin{align}
\label{eq.:alpha_1_vs_R_eq}
\frac{9 \alpha_1}{(4\pi\epsilon_0) R_{\rm eq}^7} =  \left(\frac{\mu \omega}{\hbar} \right)^2 e^{-\frac{\mu \omega}{2 \hbar} R_{\rm eq}^2}\ .  
\end{align}
This formula not only expresses a relation between $\alpha_1$ and $R_{\rm vdW} = R_{\rm eq}/2$, but also contains the QDO parameters $\mu$ and $\omega$, which are not uniquely defined for atoms. To obtain a formula connecting atomic parameters $R_{\rm vdW}$ and $\alpha_1$, we rewrite Eq.~\eqref{eq.:alpha_1_vs_R_eq} as
\begin{align}
\label{eq.:alpha_1_vs_R_vdW}
\frac{9 \alpha_1}{2^5 (4\pi\epsilon_0) R_{\rm vdW}^7} = 
\frac{e^{-(R_{\rm vdW}/\sigma_{\rm QDO})^2}}{\sigma_{\rm QDO}^4}\ ,  
\end{align}
where $\sigma_{\rm QDO} = \sqrt{\hbar/2\mu\omega}$ is the spatial variance or spread of a QDO. Within the QDO model, $\sigma_{\rm QDO}$ describes an effective atomic length, which corresponds to the Bohr radius in case of the hydrogen atom~\cite{Szabo2020}. 
According to Ref.~\cite{Fedorov2018}, the ratio
$R_{\rm vdW}/\sigma_{\rm QDO}$ decreases with increasing $\sigma_{\rm QDO}$ and the factors $\sigma_{\rm QDO}^4$ and $e^{-(R_{\rm vdW}/\sigma_{\rm QDO})^2}$ in Eq.~\eqref{eq.:alpha_1_vs_R_vdW} compensate each other. This compensation allows the QDO model to approximately capture the constant behavior of the ratio $\alpha_1/R_{\rm vdW}^7$ confirmed empirically for many atoms~\cite{Fedorov2018}. Therefore, within the QDO model, the relation between the vdW radius and the dipole polarizability can be expressed as
\begin{equation}\label{eq:Rvdwa1}
R_{\rm vdW} = A_1 (\mu \omega, R_{\rm vdW})\, \alpha_1^{\nicefrac{1}{7}}\ ,
\end{equation}
where the proportionality coefficient, as a function of the product $\mu\omega$ and the vdW radius, is given by
\begin{equation}\label{eq:r/c(dd)}
 A_1^{\mu \omega} \equiv A_1 (\mu \omega, R_{\rm vdW}) = \frac{3^{\nicefrac{2}{7}}}{2 (4\pi\epsilon_0)^{\nicefrac{1}{7}}} \left( \frac{\hbar}{\mu \omega} \, e^{ \frac{ \mu \omega R^2_{\rm vdW}}{ \hbar} } \right)^{\nicefrac{2}{7}}\, . 
\end{equation}
As was discussed in Ref.~\cite{Fedorov2018}, this coefficient can be also written in terms of the radial volume
\begin{align}
V_r = \int r^3\, n_0 (\mathbf{r})\,
d \mathbf{r} = \frac{4}{\sqrt{\pi}}
\left(\frac{\hbar}{\mu\omega}\right)^{\nicefrac 32}
\end{align}
occupied by the ground-state charge density of the QDO,
$n_0 (\mathbf{r}) \equiv |\Psi_0 (\mathbf{r})|^2 = (\frac{\mu\omega}{\pi\hbar})^{\nicefrac 32} e^{-\frac{\mu\omega}{\hbar} r^2}$, and its value at the vdW radius, $n_0 (R_{\rm vdW})$, as
\begin{align}\label{eq:A_1_mw_V}
A_1^{\mu\omega} = \frac{3^{\nicefrac{2}{7}}}{2 (4\pi\epsilon_0)^{\nicefrac{1}{7}}} \left[ \left(\frac{4}{\pi^5}\right)^{\nicefrac 13}
\frac{1}{n_0 (R_{\rm vdW}) V_r^{\nicefrac 13}} \right]^{\nicefrac 27}\ . 
\end{align}
Taking into account that $A_1$ was found to be essentially a constant for 72 atoms in the periodic table~\cite{Fedorov2018}, Eq.~\eqref{eq:A_1_mw_V} suggests a relation between an atomic volume and the electron charge density at the vdW radius. 

The results of Eqs.~\eqref{eq:r/c(dd)} and \eqref{eq:A_1_mw_V} are based on taking into account only the first term,
$V_1 \propto R^{-3}$, in the expansion of Eq.~\eqref{eq:expansion_5} for the Coulomb potential. However,
it is well-known that at least two further terms, $V_2 \propto R^{-4}$ and $V_3 \propto R^{-5}$, are important to properly describe the binding curves of vdW-bonded atomic dimers~\cite{Tang2003}. Therefore, here
we consider an extension of Eq.~\eqref{eq:Rvdwa1} by including the higher-order multipole terms from the expansion of the Coulomb potential. To this end, we evaluate the individual multipole contributions to the dispersion and exchange forces given by Eqs.~\eqref{eq:exchange_forces} and \eqref{eq:dispersion_forces}, respectively, at the equilibrium distance of vdW-bonded dimers. 
Based on the empirical findings of Ref.~\cite{Fedorov2018}, the mutual compensation of multipolar dispersion and exchange forces can ultimately be represented by the general expression
\begin{equation}
\label{eq:RvdW_alpha_l}
  R_{\rm vdW} = A_l^{\mu\omega}\, \alpha_l^{\nicefrac{2}{7(l+1)}}\ ,
\end{equation}
which extends Eq.~\eqref{eq:Rvdwa1} to the multipole polarizabilities, $\alpha_l$.
One can also rewrite Eq.~\eqref{eq:RvdW_alpha_l} in the following way
\begin{align}
\label{eq:alpha_l_vs_RvdW}
  \alpha_l = (R_{\rm vdW}/A_l^{\mu\omega})^{\nicefrac{7(l+1)}{2}}\ \ ,
\end{align}
where each multipole polarizability is expressed in terms of the vdW radius. This allows one to obtain $\alpha_l$ either from $R_{\rm vdW}$ or $\alpha_1$, for an arbitrary $l$. Since first-principles calculations of higher-order polarizabilities are computationally demanding~\cite{Lao2020}, our finding provides
an alternative way to approximate multipole polarizabilities.

Based on the expressions for the multipolar dispersion and exchange forces, we can now also explicitly calculate the proportionality coefficients $A_l^{\mu\omega} \equiv A_l (\mu \omega, R_{\rm vdW})$.
As a particularly helpful example, we can consider the force balance condition for the dipole-multipole interaction, \textit{i.e.}, the $F_{l ({\rm d} \shortminus z(l))}$-terms of Eqs.~\eqref{eq:exchange_forces} and \eqref{eq:dispersion_forces} with $z(1)={\rm d}$, $z(2)={\rm q}$, $z(3)={\rm o}$, $z(4)={\rm h}$ and $z(5)={\rm t}$.
The resulting proportionality coefficients of this series can be cast into a compact generalized formula
\begin{equation}\label{eq:r/c(dmult)}
A_l^{\mu\omega} = \left(\tfrac{D_l}{4\pi\epsilon_0}\right)^{\frac{2}{7(l+1)}} R_{\rm vdW}^{\frac{3 (l-1)}{7(l+1)}} \left( \frac{\hbar}{\mu\omega} \, e^{ \frac{ \mu \omega R^2_{\rm vdW}}{ \hbar} } \right)^{\frac{4}{7(l+1)}}\ ,
\end{equation}
with the $l$-dependent rational constants $D_l = \frac{l(l+2)(2l+1)}{2^{l+5}(l+1)}$.
Equation~\eqref{eq:r/c(dmult)} generalizes Eq.~\eqref{eq:r/c(dd)} and shows 
an additional factor of $R_{\rm vdW}^{{3 (l-1)}/{7(l+1)}}$ arising for $l > 1$. It is worth noting that the derived $A_l^{\mu\omega}$ within the QDO model formally still contain $R_{\rm vdW}$. The values for $A_l^{\mu\omega}$ calculated from Eq.~\eqref{eq:r/c(dmult)}, however, remain almost constant for any choice of realistic parameters (\textit{cf.} Table~\ref{tab:1}) as was also observed for the corresponding empirical proportionality factors~\cite{Fedorov2018}.
Moreover, in terms of the quantities related to $l=1$, the above expression can be simplified even further
\begin{align}\label{eq:r/c(dmult)_2}
A_l^{\mu\omega} = \left(\tfrac{D_l}{D_1}\right)^{\frac{2}{7(l+1)}} 
R_{\rm vdW}^{\frac{3 (l-1)}{7(l+1)}} \left(A_1^{\mu\omega}\right)^{\frac{2}{l+1}}\ .
\end{align}
The general formula given by Eq.~\eqref{eq:r/c(dmult)} allows us to obtain the proportionality coefficients $A_l^{\mu\omega}$ for every order in the dipole-multipole interactions, even without deriving further multipolar contributions to the dispersion and exchange forces.

The presented findings based on the dipole-multipole interaction can be generalized via the force balance at each order of the multipole expansion as we highlight for the quadrupole-quadrupole and octupole-octupole interactions in Appendix~\ref{appendix:B}.
Alternatively, one can use the general expression for the QDO multipole polarizabilities given by Eq.~\eqref{eq:alpha_QDO}, in order to derive $A_l^{{\rm QDO}} = R_{\rm vdW}/\alpha_{l,{\rm QDO}}^{2/7(l+1)}$ by means of Eq.~\eqref{eq:RvdW_alpha_l}. A comparison between the two approaches to the proportionality coefficients $A_l$ is given in the following section.

\subsection{Assessment of our formalism for atoms}\label{sec:num_results}

\begin{table*}[t]
\caption{Comparison between the reference ratios $A_l^{\rm ref} = R_{\rm vdW}^{\rm ref}/\alpha_{l,{\rm ref}}^{2/7(l+1)}$ for $l=\{1,2,3\}$ and their QDO counterparts $A_l^{\rm QDO} = R_{\rm vdW}^{\rm ref}/\alpha_{l,{\rm QDO}}^{2/7(l+1)}$ for $l=\{1,2,3,4,5\}$ versus the proportionality coefficients $A_l^{\mu \omega}$ for $l=\{1,2,3,4,5\}$ given by Eq.~\eqref{eq:r/c(dmult)}.
For alkali and alkaline-earth elements, we use $R_{\rm vdW}^{\rm ref}$ from the recent database of Batsanov~\cite{Batsanov2001}. For noble gas atoms and hydrogen, missing in Ref.~\cite{Batsanov2001}, the reference vdW radii are taken from Refs.~\cite{Bondi1964} and \cite{Tkatchenko2009}, respectively.
The QDO parameters $\{q, \mu, \omega \}$ are set according to
Eq.~\eqref{eq:QDOparam}, to reproduce $\alpha_1^{\rm ref}$ as well as the homoatomic dispersion coefficients ${\rm C}_6$ and ${\rm C}_8$. The three fitted quantities together with the reference quadrupole ($\alpha_2^{\rm ref}$) and octupole ($\alpha_3^{\rm ref}$) polarizabilities are taken from Refs.~\cite{Fedorov2018,Gould2016} for the noble gases (He, Ne, Ar, Kr, Xe), from Refs.~\cite{Derevianko1999, Porsev2003} for the elements in Group I (H, Li, Na, K, Rb, Cs), and from Ref.~\cite{Porsev2006} for the elements in Group II (Be, Mg, Ca, Sr, Ba). The QDO multipole polarizabilities are obtained from Eq.~\eqref{eq:alpha_QDO}, where
$\alpha_1^{\rm QDO} \equiv \alpha_1^{\rm ref}$ due to the QDO fitting procedure~\cite{Jones2013} leading to $A_1^{\rm QDO} \equiv A_1^{\rm ref}$. The average values $\langle A_l \rangle$ are calculated based on the results of the noble gas atoms. The standard deviation $\sigma = \sqrt{1/N\,\sum_X (A_l[X] - \langle A_l \rangle)^2}$ and its mean absolute relative deviation (MARD), $1/N\, \sum_X \abs{A_l[X] - \langle A_l \rangle}/ \langle A_l \rangle$, are also reported. All quantities are in atomic units (MARD in \%).}\label{tab:1}
\setlength{\tabcolsep}{6pt}
\begin{tabular}{c|c c c|c c c c c|c c c c c}
\hline\hline
Atom & $A_1^{\rm ref}$ & $A_2^{\rm ref}$ & $A_3^{\rm ref}$ & \ $A_1^{\rm QDO}$ & \ $A_2^{\rm QDO}$ & \ $A_3^{\rm QDO}$ & \ $A_4^{\rm QDO}$ & \ $A_5^{\rm QDO}$ & $A_1^{\mu \omega}$ & $A_2^{\mu \omega}$ & $A_3^{\mu \omega}$ & $A_4^{\mu \omega}$ & $A_5^{\mu \omega}$ 
\rule{0pt}{3.5ex} \\ [0.8ex]
\hline 
He   & 2.53 & 2.43 & 2.24 & 2.53 & 2.48 & 2.32 & 2.17 & 2.05 & 2.33 & 2.10 & 1.93 & 1.82 & 1.74 \\
Ne   & 2.53 & 2.44 & 2.26 & 2.53 & 2.53 & 2.38 & 2.24 & 2.12 & 2.57 & 2.27 & 2.07 & 1.94 & 1.85 \\
Ar   & 2.52 & 2.44 & 2.27 & 2.52 & 2.51 & 2.37 & 2.23 & 2.11 & 2.33 & 2.18 & 2.06 & 1.96 & 1.89 \\
Kr   & 2.55 & 2.47 & 2.29 & 2.55 & 2.55 & 2.40 & 2.26 & 2.14 & 2.35 & 2.22 & 2.10 & 2.01 & 1.94 \\
Xe   & 2.54 & 2.45 & 2.27 & 2.54 & 2.52 & 2.37 & 2.23 & 2.10 & 2.28 & 2.20 & 2.10 & 2.02 & 1.96 \\
H    & 2.50 & 2.40 & 2.19 & 2.50 & 2.43 & 2.26 & 2.11 & 2.09 & 2.06 & 1.97 & 1.88 & 1.80 & 1.75 \\
Li   & 2.40 & 2.49 & 2.33 & 2.40 & 2.48 & 2.38 & 2.26 & 1.99 & 2.50 & 2.41 & 2.29 & 2.20 & 2.14 \\
Na   & 2.53 & 2.55 & 2.40 & 2.53 & 2.56 & 2.43 & 2.30 & 2.17 & 2.50 & 2.42 & 2.32 & 2.23 & 2.17 \\
K    & 2.54 & 2.54 & 2.41 & 2.54 & 2.55 & 2.42 & 2.28 & 2.20 & 2.53 & 2.47 & 2.38 & 2.29 & 2.23 \\
Rb   & 2.61 & 2.58 & 2.46 & 2.61 & 2.60 & 2.46 & 2.31 & 2.22 & 2.58 & 2.57 & 2.42 & 2.34 & 2.27 \\
Cs   & 2.65 & 2.58 & 2.48 & 2.65 & 2.62 & 2.46 & 2.31 & 2.22 & 2.62 & 2.56 & 2.46 & 2.38 & 2.31 \\
Be   & 2.51 & 2.45 & 2.34 & 2.51 & 2.47 & 2.32 & 2.17 & 2.05 & 2.22 & 2.17 & 2.08 & 2.01 & 1.96 \\
Mg   & 2.49 & 2.42 & 2.32 & 2.49 & 2.45 & 2.29 & 2.15 & 2.03 & 2.25 & 2.21 & 2.13 & 2.07 & 2.01 \\
Ca   & 2.55 & 2.44 & 2.39 & 2.55 & 2.50 & 2.33 & 2.18 & 2.06 & 2.38 & 2.34 & 2.26 & 2.19 & 2.13 \\
Sr   & 2.61 & 2.49 & 2.43 & 2.61 & 2.55 & 2.37 & 2.22 & 2.09 & 2.45 & 2.41 & 2.32 & 2.25 & 2.19 \\
Ba   & 2.59 & 2.42 & 2.41 & 2.59 & 2.51 & 2.34 & 2.18 & 2.05 & 2.48 & 2.44 & 2.36 & 2.28 & 2.22 \\
\hline\hline 
${\langle A_l \rangle}$ & \textbf{2.54} & \textbf{2.45} & \textbf{2.27} & \textbf{2.54} & \textbf{2.52} & \textbf{2.37} & \textbf{2.23} & \textbf{2.10} & \textbf{2.37} & \textbf{2.19} & \textbf{2.05} & \textbf{1.95} & \textbf{1.88} \\
${\sigma}$ & \textbf{0.05} & \textbf{0.06} & \textbf{0.08} & \textbf{0.05} & \textbf{0.05} & \textbf{0.05} & \textbf{0.06} & \textbf{0.06} & \textbf{0.15} & \textbf{0.16} & \textbf{0.17} & \textbf{0.17} & \textbf{0.18} \\
MARD [\%] & \textbf{1.60} & \textbf{1.80} & \textbf{3.90} & \textbf{1.60} & \textbf{1.61} & \textbf{1.87} &
\textbf{2.20} & \textbf{2.47} & \textbf{5.41} & \textbf{7.13} & \textbf{8.88} & \textbf{10.14\,~} & \textbf{11.05\,~}
\end{tabular}
\end{table*}

In the previous sections, we have presented a coarse-grained approach to describe dispersion and exchange interactions between two closed-shell atoms within the QDO model.
Here, we examine the applicability of the presented formulas and apply them to analyze the ratio between the vdW radius and multipole
polarizabilities for atoms, thus demonstrating the validity of the scaling law of Eq.~\eqref{eq:RvdW_alpha_l} obtained within the QDO model.
Our analysis will be focused on hydrogen,
noble gases from He to Xe, alkali atoms from Li to Cs, and alkaline-earth elements from Be to Ba. 
To this end, the atomic multipole polarizabilities, $\alpha_l$, are either taken from high-level \emph{ab initio} calculations in the literature~\cite{Derevianko1999, Porsev2003,Porsev2006}, $\alpha_l^{\rm ref}$, or calculated by means of Eq.~\eqref{eq:alpha_QDO}, $\alpha_l^{\rm QDO}$. We determine $q$, $\mu$ and $\omega$ for each atom by means of Eq.~\eqref{eq:QDOparam}, using accurate reference data for the set $\lbrace \alpha_1^{\rm ref}, {\rm C}_6, {\rm C}_8 \rbrace$~\cite{Fedorov2018,Gould2016,Derevianko1999, Porsev2003,Porsev2006}, as explained in Section~\ref{sec:method}. Here, due to the fact that the QDO parameters are set to reproduce the dipole polarizability,
we have $\alpha_1^{\rm QDO} \equiv \alpha_1^{\rm ref}$, for all considered atoms.

While eventually it would be interesting and important to extend our analysis to a broader set of atoms and small molecules, we are not aware of a comprehensive set of accurate data for atomic and molecular multipole polarizabilities. Accurate \textit{ab initio} reference calculations of $\alpha_l$ in general require demanding computational approaches with sophisticated treatment of electron correlation effects and, especially with increasing order $l$, large and diffuse basis sets~\cite{Woon1994,Lao2018,Lao2020}. As a result, calculating converged multipolar polarizabilities is difficult, a problem which is further enhanced by the numerical aspects associated with finite-field derivative techniques as used in such calculations.
Experimental determination, on the other hand, is subject to origin and orientational dependencies as well as a strong influence of thermal effects~\cite{Bishop1990,Kuemmel2000,Lao2020}, which can introduce considerable uncertainties --- in particular with increasing system size or multipole order.

To apply the derived formulas, a set of reference
vdW radii, $R_{\rm vdW}^{\rm ref}$, is required. In the case of alkali as well as alkaline-earth elements, these radii are taken from the recent database of Batsanov~\cite{Batsanov2001}. For noble-gas atoms, missing in Ref.~\cite{Batsanov2001}, we use the database of Bondi~\cite{Bondi1964}, which provides often used values of vdW radii for Group 18 of the periodic table. In addition, for hydrogen, we use $R_{\rm vdW}^{\rm ref} = 3.1$ a.u. from Ref.~\cite{Tkatchenko2009}, where it was theoretically estimated based on the atomic charge density. This value was shown to work well for the relation between the atomic dipole polarizability and vdW radius in Ref.~\cite{Fedorov2018}.

Both, the vdW radii of Batsanov and Bondi, are extracted from experimental crystallographic structural data. However, it is important to mention that a straightforward definition of the vdW radius is only possible for noble-gas atoms as inert elements with closed valence shells. For other atoms, $R_{\rm vdW}$ is evaluated by considering a variety of different molecular crystal structures and extracting neighboring atom--atom distances, where each atom belongs to a different closed-shell molecule. This definition is especially subtle for chemical elements with spin-polarized valence shells, such as alkali atoms, which can form bonds with different spin states. Therefore, one has to keep in mind that existing vdW radii are just statistical quantities for most chemical elements.

\begin{figure}[t]
 \includegraphics[width=0.48 \textwidth]{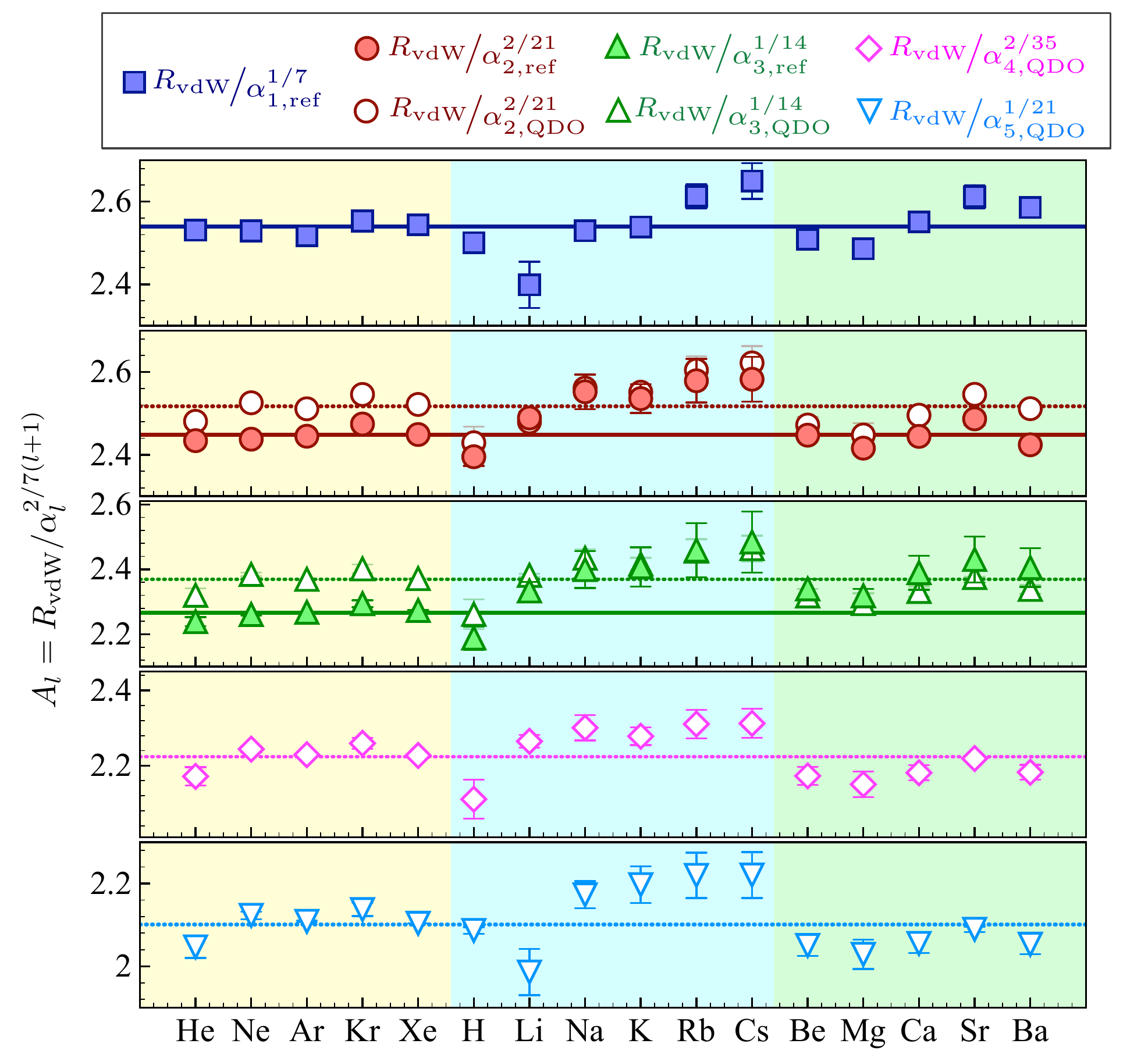}
 \caption{The ratios $R_{\rm vdW}/\alpha_{l,{\rm ref}}^{2/7(l+1)}$ listed in Table~\ref{tab:1} are shown (filled symbols) with respect to the ``universal'' values $A_1 = 2.54$~a.u., $A_2 = 2.45$~a.u. and $A_3 = 2.27$~a.u., represented with blue, red and green solid lines, respectively. By contrast, the ratios $R_{\rm vdW}/\alpha_{l,{\rm QDO}}^{2/7(l+1)}$ (open symbols) are plotted with respect to the constant values obtained from the QDO polarizabilities, $A_{2}^{{\rm QDO}} = 2.52$~a.u., $A_{3}^{{\rm QDO}} = 2.37$~a.u., $A_{4}^{{\rm QDO}} = 2.23$~a.u. and $A_{5}^{{\rm QDO}} = 2.10$~a.u., shown in red, green, fuchsia and light blue dashed lines, respectively. The noble gases are shown in the yellow box, the elements of the Group~I in the light blue box and the elements of the Group~II in the light green box. The error bars represent the relative errors ${\rm R.E.} = \abs{A_l[X] - A_l}/A_l \,$ of each species $X$, where $A_l$ are the ``universal'' values expressed in Eqs.~\eqref{eq:Rvdw_a} for the ratios $R_{\rm vdW}/\alpha_{l,{\rm ref}}^{2/7(l+1)}$ and in Eqs.~\eqref{eq:Rvdw_aQDO} for $R_{\rm vdW}/\alpha_{l,{\rm QDO}}^{2/7(l+1)}$.} \label{fig:2}
\end{figure}

First, we analyze the empirical proportionality constants
\begin{equation}
  A_l^{\rm ref} = R_{\rm vdW}^{{\rm ref}} / \alpha_{l,{\rm ref}}^{\nicefrac{2}{7(l+1)}}
  \label{eq:A_l-ref}
\end{equation}
based on the reference data of $R_{\rm vdW}$ and
the atomic dipole ($\alpha_1^{{\rm ref}}$), quadrupole ($\alpha_2^{{\rm ref}}$) and octupole ($\alpha_3^{{\rm ref}}$) polarizabilities~\cite{Fedorov2018,Gould2016,Derevianko1999, Porsev2003,Porsev2006}. The results are shown in Table~\ref{tab:1}, for the chosen test set of 16 chemical elements.

For noble gases, $A_l^{\rm ref}$ is essentially constant, and, in line with Ref.~\cite{Fedorov2018}, we find $\langle A_1^{\rm ref} \rangle = 2.54$~a.u.
Moreover, this result is further specified in form of the unified formula, $R_{\rm vdW} (\alpha_1) = (a_0^4/4\pi\epsilon_0)^{1/7}\,(1/\alpha_f)^{4/21}\,\alpha_1^{1/7}$ with $a_0$ and $\alpha_f$ denoting the Bohr radius and the fine-structure constant, respectively~\cite{Tkatchenko2020}. Using Hartree atomic units, this can be further simplified to $R_{\rm vdW} (\alpha_1) \approx (137)^{4/21}\,\alpha_1^{1/7}$, which gives an appropriate proportionality factor of $2.55$~a.u., in excellent agreement with the empirical value of $\langle A_1^{\rm ref} \rangle$.
For the higher-order multipoles, we obtain $\langle A_2^{\rm ref} \rangle = 2.45$~a.u. and $\langle A_3^{\rm ref} \rangle = 2.27$~a.u..
The resulting values for $A_l^{\rm ref}$ remain close to the average values determined for noble-gas atoms for all the atoms shown in Table~\ref{tab:1}, 
where as a general trend alkali metal atoms show the largest deviation with an average relative deviation of 4.3\,\% (compared to 2.4\,\% for alkaline earth metals and 0.5\,\% for noble gases).
The alkali atoms possess a relatively weakly bound and, therefore, highly polarizable single valence electron. This feature and the different possible spin states of alkali atoms in molecular solid-state systems arguably allow the vdW radii observed for alkali metals to change widely, causing the largest deviation for the empirical constants $A_l^{\rm ref}$ from their average values, as also illustrated in Fig.~\ref{fig:2}.

Overall, the observed deviations in $A_l^{\rm ref}$ stay within 4\,\% ($\hat{=}$ 0.1~a.u.) for all considered species except for Cs and Rb, which show a slightly higher average relative deviation of 6.3\,\% and 5.5\,\%, respectively.
Hence, we suggest that
the coefficients $A_l^{\rm ref}$ can be considered as universal constants for the studied atomic species.
Taking the average values $A_l = \langle A_l^{\rm ref} \rangle$ for the noble-gas atoms we can thus write the unified relations between the vdW radius and the dipole, quadrupole and octupole polarizabilities,
\begin{equation}
\begin{alignedat}{3}
R_{\rm vdW}(\alpha_1) &= A_1 \, \alpha_1^{\nicefrac{1}{7}}\ \ &&,\ \ \ \ A_1 &&= 2.54\ {\rm a.u.} \label{eq:Rvdw_a} \\
R_{\rm vdW}(\alpha_2) &= A_2 \, \alpha_2^{\nicefrac{2}{21}}\ \ &&,\ \ \ \ A_2 &&= 2.45\ {\rm a.u.} \\
R_{\rm vdW}(\alpha_3) &= A_3 \, \alpha_3^{\nicefrac{1}{14}}\ \ &&,\ \ \ \ A_3 &&= 2.27\ {\rm a.u.} 
\end{alignedat}
\end{equation}
which are equivalent to the empirical relations reported in Ref.~\cite{Fedorov2018}. 
The relations obtained above can be used for at least three different purposes. First, the vdW radius of atoms can now be calculated given any single multipolar atomic polarizability. This polarizability can correspond to a free atom or an atom in a molecule or material~\cite{Tkatchenko2009}. The vdW radius can then be used for a conceptual understanding of an atom in its environment or for practical calculations of the vdW energy~\cite{Tkatchenko2009,Hermann2020}. Second, given the dipole polarizability, one can accurately determine multipole polarizabilities (at least up to octupole) from Eq.~\eqref{eq:Rvdw_a}. In fact, this approach is substantially more reliable for atoms than using the QDO model for multipole polarizabilities. A third application would be the possibility to determine atomic multipole polarizabilities from calculated or measured atomic vdW radii. A potential downside of this application is that a small error in the vdW radius would result in a large error for multipole polarizabilities,
according to Eq.~\eqref{eq:Rvdw_a}.

\begin{figure}[t]
 \includegraphics[width=0.48 \textwidth]{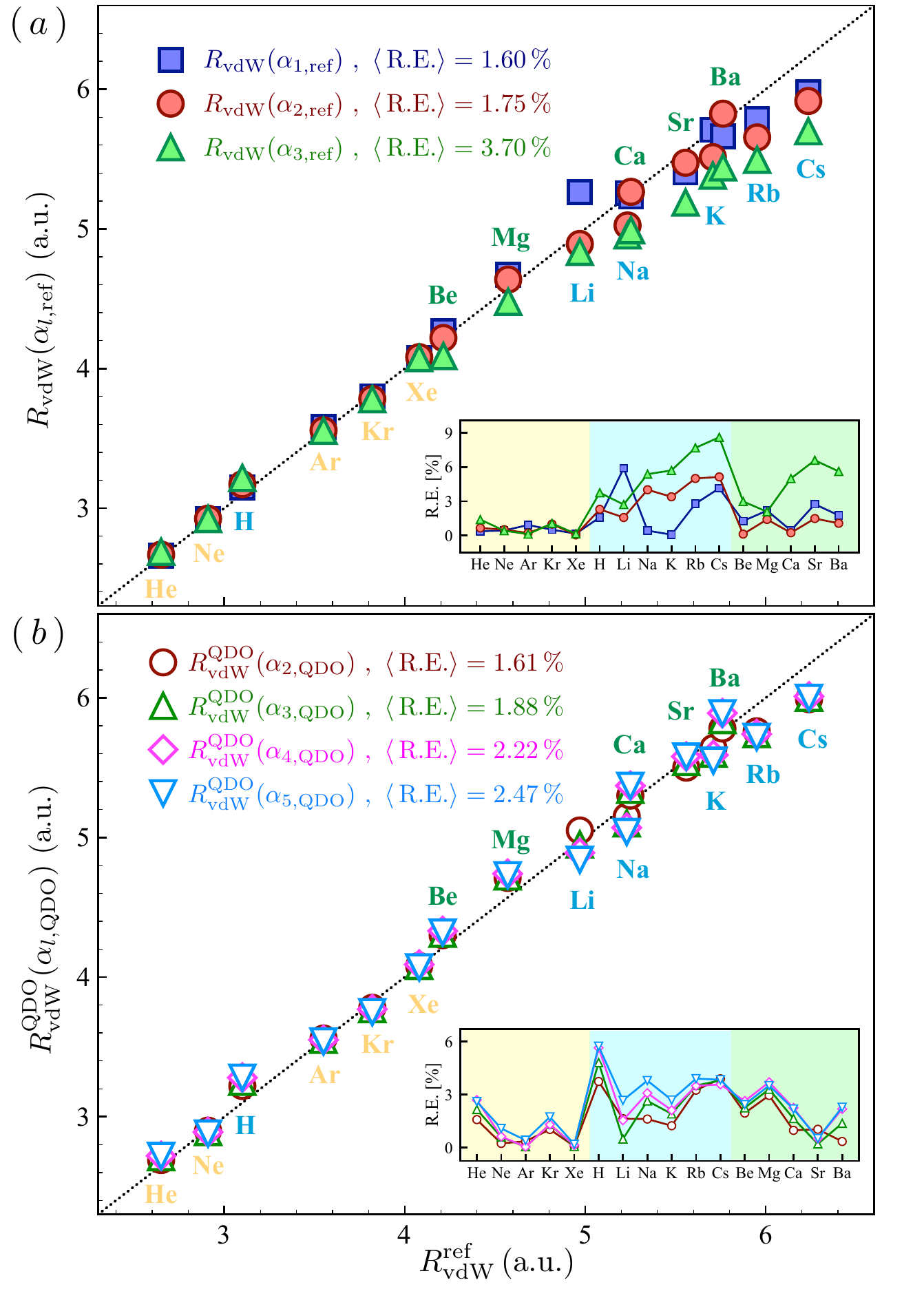}
 \caption{The vdW radius as a function of the multipole polarizabilities, $R_{\rm vdW} (\alpha_l) = A_l\, \alpha_l^{\nicefrac{2}{7(l+1)}}$, obtained by means of 
 (a) Eqs.~\eqref{eq:Rvdw_a} and (b) Eqs.~\eqref{eq:Rvdw_aQDO}
 is presented, taking into account that $R_{\rm vdW}^{\rm QDO}(\alpha_{1,{\rm QDO}}) \equiv R_{\rm vdW} (\alpha_{1,{\rm ref}})$, in comparison to the reference values 
 $R_{\rm vdW}^{{\rm ref}}$~\cite{Tkatchenko2009,Batsanov2001,Bondi1964} for three classes of species: noble gases (in yellow), H + alkali-metal atoms (Group~I, in blue) and alkaline-earth atoms (Group~II, in green). The vdW radii are shown (a) for the reference~\cite{Jones2013, Derevianko1999, Porsev2003, Porsev2006} multipole polarizabilities $\alpha_{1,{\rm ref}}$, $\alpha_{2,{\rm ref}}$ and $\alpha_{3,{\rm ref}}$ (blue, red and green filled symbols) and (b) for the QDO multipole polarizabilities $\alpha_{2,{\rm QDO}}$, $\alpha_{3,{\rm QDO}}$, $\alpha_{4,{\rm QDO}}$ and $\alpha_{5,{\rm QDO}}$ given by Eq.~\eqref{eq:alpha_QDO} (red, green, fuchsia and light blue open symbols). The insets show the relative errors, ${\rm R.E.} = \abs{R_{\rm vdW}(\alpha_1) - R_{\rm vdW}^{{\rm ref}}}/R_{\rm vdW}^{{\rm ref}}$, for the noble gases (yellow box), the alkali-metals + hydrogen (light blue box) and the alkaline-earth elements (light green box). In addition, the mean values of the relative errors, $\langle {\rm R.E.} \rangle$, are reported in the legens, for each considered multipole order.} \label{fig:3}
\end{figure}

The quantity $R_{\rm vdW} (\alpha_l)$ defined in Eq.~\eqref{eq:Rvdw_a} represents an effective vdW radius expressed in terms of the multipolar polarizability. To demonstrate the validity of this new definition for the vdW radius, Fig.~\ref{fig:3}(a) compares the results for $R_{\rm vdW} (\alpha_l) = A_l\, \alpha_l^{2/7(l+1)}$ with
$l = \{1,2,3\}$ to the reference values, $R_{\rm vdW}^{{\rm ref}}$. There is an excellent correlation between $R_{\rm vdW}(\alpha_1)$ and its reference value for all considered elements, with a maximum relative error ${\rm R.E.} = \abs{R_{\rm vdW}(\alpha_1) - R_{\rm vdW}^{{\rm ref}}}/R_{\rm vdW}^{{\rm ref}}$ of 0.91\,\% for the noble gases, 2.74\,\% for alkaline-earth (Group II) elements and 5.90\,\% for hydrogen and alkali metals (Group I). 
The increasing errors when going from alkaline-earth to alkali metals are related to the increase in the statistical errors of $R_{\rm vdW}^{\rm ref}$ stemming from the increasingly complicated evaluation of the vdW radii based on experimental crystal-structure data. 
Indeed, this evaluation becomes less accurate for elements with more pronounced metallic properties~\cite{Batsanov2012,Batsanov2001}. Comparing Groups I and II of the periodic table, the statistical errors in $R_{\rm vdW}^{\rm ref}$ of the alkaline-earth elements are smaller since they have closed $s$-electron shell, which makes their behavior closer to that of the noble gases with completely closed valence shells. 

Although the dipole polarizability $\alpha_{1,{\rm ref}}$ is known with high accuracy for many chemical elements in the periodic table, the accurate determination of higher-order multipolar polarizabilities is more involved. Indeed, Fig.~\ref{fig:3}(a) shows an increase in the maximum R.E. for $R_{\rm vdW}(\alpha_2)$ (within 0.99\,\% for noble gases, 1.48\,\% for Group I and 5.14\,\% for Group II) and, subsequently, for $R_{\rm vdW}(\alpha_3)$ (within 1.39\,\% for noble gases, 6.60\,\% for Group I and $8.62$\,\% for Group II). This analysis is validated in Fig.~\ref{fig:2} as well, where we compare the average values $A_1 = 2.54$~a.u., $A_2 = 2.45$~a.u. and $A_3 = 2.27$~a.u. with the relative ratios $R_{\rm vdW}/\alpha_l^{2/7(l+1)}$, for $l=\{1,2,3\}$.

It is also noteworthy that reference values for higher-order polarizabilities are rather limited in literature, with the exception of hydrogen, the only element in the periodic table for which the multipole polarizability $\alpha_l^{\rm H}$ is known analytically~\cite{Kharchenko2015}. 
Therefore, we employ the multipole polarizabilities obtained within the QDO model by means of Eq.~\eqref{eq:alpha_QDO}. Notably, the QDO coefficients $A_l^{{\rm QDO}} = R_{\rm vdW}/\alpha_{l,{\rm QDO}}^{2/7(l+1)}$, for $l = \{1, 2, 3, 4 ,5\}$, are practically constant for all noble-gas atoms, which leads to the QDO set of relations
\begin{equation}
\begin{aligned}
R_{\rm vdW}^{\rm QDO}(\alpha_1) &= \langle A_1^{{\rm QDO}} \rangle \, \alpha_{1}^{\nicefrac{1}{7~}}\ \ , \ \ \langle A_1^{{\rm QDO}} \rangle = 2.54\ {\rm a.u.} \label{eq:Rvdw_aQDO} \\
R_{\rm vdW}^{\rm QDO}(\alpha_2) &= \langle A_2^{{\rm QDO}} \rangle \, \alpha_{2}^{\nicefrac{2}{21}}\ \ , \ \ \langle A_2^{{\rm QDO}} \rangle = 2.52\ {\rm a.u.} \\
R_{\rm vdW}^{\rm QDO}(\alpha_3) &= \langle A_3^{{\rm QDO}} \rangle \, \alpha_{3}^{\nicefrac{1}{14}}\ \ , \ \ \langle A_3^{{\rm QDO}} \rangle = 2.37\ {\rm a.u.} \\
R_{\rm vdW}^{\rm QDO}(\alpha_4) &= \langle A_4^{{\rm QDO}} \rangle \, \alpha_{4}^{\nicefrac{2}{35}}\ \ , \ \ \langle A_4^{{\rm QDO}} \rangle = 2.23\ {\rm a.u.} \\
R_{\rm vdW}^{\rm QDO}(\alpha_5) &= \langle A_5^{{\rm QDO}} \rangle \, \alpha_{5}^{\nicefrac{1}{21}}\ \ , \ \ \langle A_5^{{\rm QDO}} \rangle = 2.10\ {\rm a.u.} 
\end{aligned}
\end{equation}
These results are shown in Fig.~\ref{fig:3}(b), where a good agreement between $R_{\rm vdW}^{\rm QDO}(\alpha_l)$ and $R_{\rm vdW}^{\rm ref}$ is observed for $l=\{2,3,4,5\}$, in addition to the case of $l=1$ for which we have $R_{\rm vdW}^{\rm QDO}(\alpha_{1,{\rm QDO}}) \equiv R_{\rm vdW} (\alpha_{1,{\rm ref}})$.
We note that the QDO model is constructed, by definition, on the dispersion coefficients and the QDO polarizabilities (with $l > 1$) are underestimated for the noble-gas atoms with respect to the reference data~\cite{Jones2013}.
Consequently, the QDO proportionality coefficients $\langle A_2^{{\rm QDO}} \rangle = 2.52$~a.u. and $\langle A_3^{{\rm QDO}} \rangle = 2.37$~a.u. are overestimated with respect to the determined ``universal'' values $A_2 = 2.45$~a.u. and $A_3 = 2.27$~a.u., as also shown in Fig.~\ref{fig:2}. Therefore, one can expect that the higher-order QDO coefficients, $\langle A_4^{{\rm QDO}} \rangle = 2.23$~a.u. and $\langle A_5^{{\rm QDO}} \rangle = 2.10$~a.u., are also overestimated.

To further assess the scaling law of Eq.~\eqref{eq:RvdW_alpha_l}, we compare the resulting empirical constants with the proportionality coefficients $A_l^{\mu \omega}$, obtained by means of Eq.~\eqref{eq:r/c(dmult)}.
Table~\ref{tab:1} summarizes the results for $A_l^{\mu \omega}$ compared with $A_l^{\rm ref}$ and $A_l^{\rm QDO}$. Considering the noble-gas atoms only, we find the following averaged values
\begin{align}
& \langle A_1^{\mu \omega} \rangle = \nonumber 2.37\ {\rm a.u.},
\ \langle A_2^{\mu \omega} \rangle = 2.19\ {\rm a.u.},
\ \langle A_3^{\mu \omega} \rangle = 2.05\ {\rm a.u.}
\\ & \langle A_4^{\mu \omega} \rangle = 1.95\ {\rm a.u.},
\ \langle A_5^{\mu \omega} \rangle = 1.88\ {\rm a.u.}\ \ ,
\end{align}
which shall serve as the suggested values of $A_l^{\mu \omega}$. The mean absolute relative deviations from this reference across all considered elements are 0.15~a.u.~(5.41\,\%) for the dipole-dipole term, 
0.16~a.u.~(7.13\,\%) for the dipole-quadrupole term, 0.17~a.u.~(8.88\,\%) for the dipole-octupole term, 0.17~a.u.~(10.14\,\%) for the dipole-hexadecapole term and 0.18~a.u.~(11.05\,\%) for the dipole-triakontadipole term in the multipole expansion. Hence, $A_l^{\mu\omega}$ do not remain constant among all considered elements, 
in contrast to the respective atomic proportionality constants $A_l^{\rm ref}$. 
The deviations between the two sets of coefficients amount to $0.17$~a.u.~(6.6\,\%) for $A_1$, $0.25$~a.u.~(10.3\,\%) for $A_2$, and $0.21$~a.u.~(9.4\,\%) for $A_3$, where the derived $A_l^{\mu\omega}$ are always lower than the empirical reference values $A_l^{\rm ref}$.
In both cases, we consistently observe a decrease of the average values and an increase in the standard deviations with increasing multipole order. 
Interestingly, the standard deviations among $A_l^{\rm ref}$
and $A_l^{\rm QDO}$ are comparable to each other but considerably smaller than $\sigma$ shown in Table~\ref{tab:1} for the corresponding $A_l^{\mu\omega}$.
Hence, the simplifications in the coarse-grained description of valence electrons within the QDO model
and its parametrization lead to a less accurate determination of the proportionality coefficients, based on Eq.~\eqref{eq:r/c(dmult)}, than their indirect evaluation based on the QDO multipole polarizabilities calculated by means of Eq.~\eqref{eq:alpha_QDO}. However, even 
$A_l^{\rm QDO}$ obtained for noble gases still show noticeable deviations from a constant behavior. It is unclear yet what aspect of atoms makes $A_l^{\rm ref}$ behave as ``universal'' constants for different chemical elements. Nevertheless, we expect the clarification of this question to be crucial for the eventual improvement of the QDO model.

\begin{figure}[t]
 \includegraphics[width=0.48 \textwidth]{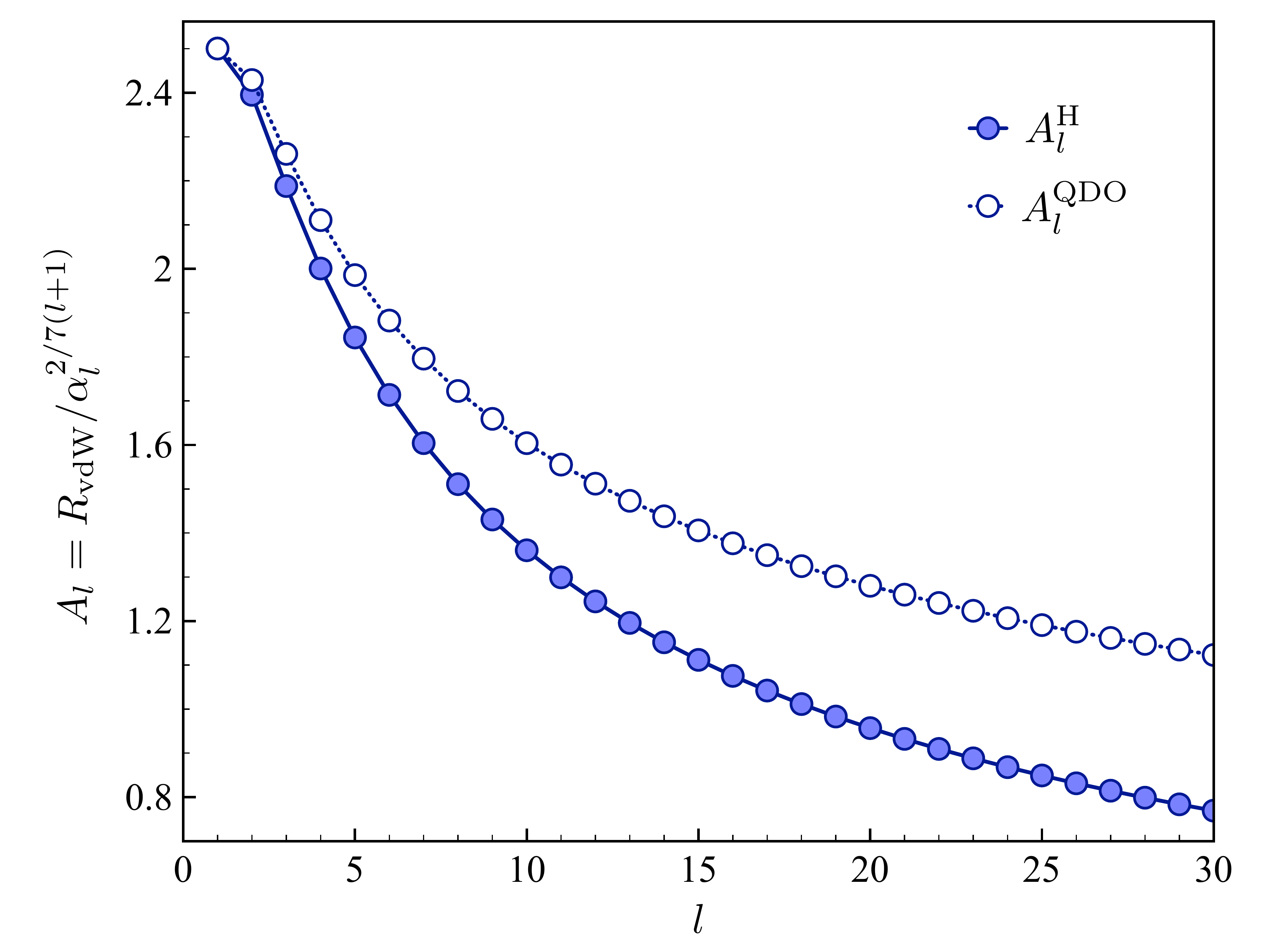}
 \caption{The coefficients $A_l = R_{\rm vdW}/\alpha_l^{2/7(l+1)}$ for hydrogen: the analytical solution $A_l^{{\rm H}}$, from the multipole polarizabilities $\alpha_l^{{\rm H}} = (4 \pi \epsilon_0) a_0^{2l+1}  (2l+1)! (l+2)/2^{2l} l$~\cite{Kharchenko2015}, where $a_0$ is the Bohr radius, is shown (blue filled circles) in comparison to the results of the QDO model, $A_l^{{\rm QDO}}$ (blue open circles), which are obtained from $\alpha_l^{{\rm QDO}}$ calculated by means of Eq.~\eqref{eq:alpha_QDO}.} 
\label{fig:4}
\end{figure}

Finally, we calculate the ratio $R_{\rm vdW}/\alpha_l^{2/7(l+1)}$ for the hydrogen atom, taking into account the known analytical expression of its multipole polarizabilities given in Ref.~\cite{Kharchenko2015} as $\alpha_l^{{\rm H}} = (4 \pi \epsilon_0)\, a_0^{2l+1} (2l+1)! (l+2)/2^{2l} l$, and compare it with the proportionality coefficients $A_l^{\rm QDO}$, which are obtained from $\alpha_{l,{\rm QDO}}$ provided by Eq.~\eqref{eq:alpha_QDO}. 
As shown in Fig.~\ref{fig:4}, the current QDO parametrization yields quite accurate results for dipole, quadrupole and octupole orders, but then exhibits an increasing overestimation of the proportionality constant with increasing multipole rank $l$. Given that the vdW radius is fixed at its reference value~\cite{Tkatchenko2009}, this reflects the fact that the QDO model predicts underestimated multipolar polarizabilities~\cite{Jones2013}. In order to improve the model, a possible future step is to use highly accurate theoretical or experimental reference data
for the quadrupole polarizability instead of the dispersion coefficient ${\rm C}_8$ in the parametrization scheme, 
which would increase the accuracy of the higher-order
$\alpha_{l,\mathrm{QDO}}$ values. Consequently, such new parametrization would yield smaller values of $A_l^{\rm QDO}$, providing better agreement with the reference data and improving the relations between multipole polarizabilities and the equilibrium distance in vdW-bonded atomic dimers. Figure~\ref{fig:4} also shows that the $A_l^{\rm QDO}$ curve produces a slight deviation in the general trend of the reference proportionality coefficients at $l=2$. This kink stems from the nonmonotonic behavior of the multipole contributions to the exchange energy, where $J^{{\rm ex}}_{2({\rm d \shortminus q})}$ is larger than $J^{{\rm ex}}_{1({\rm d \shortminus d})}$ (see our discussion at the end of the subsection B).

One possible explanation for the difference between the polarizabilities of the QDO model and the hydrogen atom is the contribution of excitations to continuum states in the latter case. The QDO model has no continuum states and can only effectively describe such excitations. Despite the observed deviations in the  higher-order multipole polarizabilities and the corresponding proportionality coefficients, the QDO approach allowed us to verify the scaling relation $R_{\rm vdW} \propto \alpha_l^{2/7(l+1)}$, which is remarkably valid for atoms. 

\section{Discussion and Summary}
We have presented a coarse-grained description of the repulsive force due to the Pauli principle and attractive force due to dispersive fluctuations between two closed-shell atoms or molecules. Our formalism is based on two interacting quantum Drude oscillators, for which the dispersion and exchange-repulsion energies up to an arbitrary order in the multipole expansion of the Coulomb potential were derived. The obtained formulas can be employed for constructing and rationalizing effective interaction potential models, as well as for finding new scaling laws between electronic and geometric properties of atoms and molecules.

As a practical illustration of our theory, we investigated a mutual compensation between the repulsive exchange and attractive dispersion forces for each term in the multipole expansion. The results confirm and extend the recently proposed relations~\cite{Fedorov2018} between atomic multipole polarizabilities, $\alpha_l$, and the van der Waals radius, $R_{\rm vdW}$. The 
generalized scaling law, $R_{\rm vdW} = A_l\,\alpha_l^{2/7(l+1)}$, is compelling because it connects an electronic response property of a single atom (atomic multipole polarizability) with the equilibrium distance in a homonuclear dimer. 

Let us enumerate some of the potential applications of the formulas presented in this paper and possible future research directions:
\begin{itemize}
    \item First and foremost, the relation between atomic vdW radius and atomic polarizabilities, $R_{\rm vdW} = A_l\,\alpha_l^{2/7(l+1)}$, dispenses with the need to indirectly measure vdW radii. Once the polarizability is calculated for a free atom or an atom in a molecule/material, the vdW radius can be computed from the formula above. Subsequently, the vdW radius can be used as a proxy for an atomic size, as an effective radius in interatomic vdW potentials or in damping functions for vdW-inclusive electronic-structure calculations.
    We remark that in quantum mechanics, many possible definitions can be made for an effective atomic size. Our derivations provide a novel definition of the atomic vdW radius in terms of observable quantities --- atomic multipole polarizabilities. Obviously, a more detailed comparison of calculated vdW radii to experiment would be welcome by measuring effective vdW radii in a wide set of systems and comparing the measured radii to first-principles calculations and our formulas.
    
    \item Our formula allows a straightforward and accurate calculation of atomic multipole polarizabilities from the dipole polarizability. Given $\alpha_1$ and the universal values of $A_l^{\rm{ref}}$ obtained in this work, any multipole polarizability can now be calculated as a function of these two parameters. This is especially important given the high computational cost of calculating multipole polarizabilities from first principles of quantum mechanics. 
    Going further, it would be interesting to assess different recursive relations between $\alpha_l$ and $\alpha_{l+1}$ polarizabilities based on the QDO model and the definition of the vdW radius. 
    
    \item Our analytical results allow calculating multipole polarizabilities $\alpha_l$ for an arbitrary value of $l$. Such data become increasingly important in coarse-grained models, which describe molecular response by increasingly larger fragments. For example, one might want to describe the response of a protein, where one QDO models the response of each amino acid. Similar to electrostatics, where higher multipoles become of growing importance when increasing the fragment size, polarization response follows the same trend. Hence, we expect our formulas to play a key role in the development of coarse-grained models for chemical and biological matter.  
    
    \item While most of the results in this paper were presented for homonuclear dimers, an accurate combination rule is already known for computing equilibrium distances, $R_{\rm eq}^{\rm AB}$, in heteronuclear dimers~\cite{Fedorov2018}
    \begin{align}
    R_{\rm eq}^{\rm AB} = 2 \times A_l [(\alpha_{\rm A} + \alpha_{\rm B})/2 ]^{\nicefrac{2}{7(l+1)}} \quad .
    \end{align}
    This formula allows the calculation of equilibrium distances in heteronuclear closed-shell dimers solely based on the knowledge of atomic polarizabilities of each atom. The derivation of such combination rules from first principles requires generalizing the Pauli principle to QDOs with different parameters and will be a subject of future work.
    
    \item The determination of $R_{\rm eq}^{\rm AB}$ for two atoms A and B from their dipole polarizabilities provides a way to construct generalized Tang-Toennies-type potentials~\cite{Tang1995,Tang2003,Sheng2020,Tang1998} that require only one adjustable parameter: the equilibrium interaction energy $E_{\rm eq}$. It remains to be investigated whether the asymptotic dispersion coefficients could be connected to $E_{\rm eq}$, allowing one to construct parameter-free Tang-Toennies-type interatomic potentials for closed-shell systems.
    
    \item Last but not least, the relation between $R_{\rm vdW}$ and the polarizability could be used to develop a more general and more accurate parametrization of the QDO model. Namely, the universality of $A_l^{\rm ref}$ coefficients holds for atoms but it is not such a good approximation within the QDO model itself. One could enforce the obtained relation, $R_{\rm vdW} = A_l\,\alpha_l^{2/7(l+1)}$ using ``universal'' values $A_l$, to hold on average for the QDO model during the parametrization procedure. This is a direction of our current study.   
    
\end{itemize}

Ultimately, the close connection between vdW attraction and Pauli repulsion unveiled in our work paves the way for the construction of efficient coarse-grained models for the description of the exchange-repulsion interaction in atomic and molecular systems. Together with the well established success of the QDO model in describing vdW dispersion, our results also provide the basis for constructing consistent and minimally-empirical models for interatomic and intermolecular forces.\\
 
\section*{ACKNOWLEDGMENTS}
The authors acknowledge financial support from the European Research Council via the ERC Consolidator Grant \lq\lq BeStMo (725291)\rq\rq as well as the Luxembourg National Research Fund via the FNR CORE Jr project \lq\lq PINTA(C17/MS/11686718)\rq\rq\ and the AFR PhD Grant \lq\lq CNDTEC(11274975)\rq\rq.

\appendix 

\section{Multipole Expansion of Coulomb Potential}\label{appendix:A}

Here, we present the contributions $V_n$ to the multipole expansion of the Coulomb potential given by Eq.~\eqref{eq:expansion_5}.

{\tiny
\begin{align*}
 V_{2({\rm d \shortminus q})} &= \tfrac{3 q_A q_B}{2 R^4} \big({{-}}5 (\textbf{r}_1{{\cdot}}\textbf{R}_0)^2 \textbf{r}_2{{\cdot}}\textbf{R}_0 {{+}} (\textbf{r}_1^2 {{-}} 2 \textbf{r}_1{{\cdot}}\textbf{r}_2) \textbf{r}_2{{\cdot}}\textbf{R}_0 {{+}} \textbf{r}_1{{\cdot}}\textbf{R}_0 ({{-}}\textbf{r}_2^2 {{+}} 2 \textbf{r}_1{{\cdot}}\textbf{r}_2 \\ 
 & \, {{+}} 5 (\textbf{r}_2{{\cdot}}\textbf{R}_0)^2) \big)  \\
 V_{3({\rm d \shortminus o})} &= \tfrac{q_A q_B}{2 R^5} \big( 5 \textbf{r}_1{{\cdot}}\textbf{R}_0 \textbf{r}_2{{\cdot}}\textbf{R}_0 (3 (\textbf{r}_1^2 {{+}} \textbf{r}_2^2) {{-}} 7 (\textbf{r}_1{{\cdot}}\textbf{R}_0)^2 {{-}} 7 (\textbf{r}_2{{\cdot}}\textbf{R}_0)^2) {{-}} 3 \textbf{r}_1{{\cdot}}\textbf{r}_2 (\textbf{r}_1^2 \\ 
 & \, {{+}} \textbf{r}_2^2 {{-}} 5 (\textbf{r}_1{{\cdot}}\textbf{R}_0)^2 {{-}} 5 (\textbf{r}_2{{\cdot}}\textbf{R}_0)^2) \big)  \\
 V_{3({\rm q \shortminus q})} &= \tfrac{3 q_A q_B}{4 R^5} \big(2 (\textbf{r}_1{{\cdot}}\textbf{r}_2)^2 {{-}} 20 \textbf{r}_1{{\cdot}}\textbf{R}_0 \textbf{r}_1{{\cdot}}\textbf{r}_2 \textbf{r}_2{{\cdot}}\textbf{R}_0 {{-}} 
 5 (\textbf{r}_1{{\cdot}}\textbf{R}_0)^2 (\textbf{r}_2^2 {{-}} 7 (\textbf{r}_2{{\cdot}}\textbf{R}_0)^2) \\ 
 & \, {{+}} \textbf{r}_1^2 (\textbf{r}_2^2 {{-}} 5 (\textbf{r}_2{{\cdot}}\textbf{R}_0)^2) \big)  \\
 V_{4({\rm d \shortminus h})} &= \tfrac{5 q_A q_B}{8 R^6} \big( 28 (\textbf{r}_1{{\cdot}} \textbf{R}_0)^3 \textbf{r}_1{{\cdot}} \textbf{r}_2 {{+}} 42 \textbf{r}_1^2 (\textbf{r}_1{{\cdot}} \textbf{R}_0)^2 \textbf{r}_2{{\cdot}} \textbf{R}_0 {{-}} 
  63 (\textbf{r}_1{{\cdot}} \textbf{R}_0)^4 \textbf{r}_2{{\cdot}} \textbf{R}_0 \\ 
 & \, {{+}} 
  3 \textbf{r}_1{{\cdot}} \textbf{R}_0 (\textbf{r}_2^4 {{-}} 4 \textbf{r}_1^2 \textbf{r}_1{{\cdot}} \textbf{r}_2 {{-}} 14 \textbf{r}_2^2 (\textbf{r}_2{{\cdot}} \textbf{R}_0)^2 {{+}} 21 (\textbf{r}_2{{\cdot}} \textbf{R}_0)^4) {{+}} 
  \textbf{r}_2{{\cdot}} \textbf{R}_0 ({{-}}3 \textbf{r}_1^4 \\ 
 & \, {{+}} 4 \textbf{r}_1{{\cdot}} \textbf{r}_2 (3 \textbf{r}_2^2 {{-}} 7 (\textbf{r}_2{{\cdot}} \textbf{R}_0)^2)) \big) \\
 V_{4({\rm q \shortminus o})} &= \tfrac{5 q_A q_B}{4 R^6} \big( {{-}}7 (\textbf{r}_1{{\cdot}} \textbf{R}_0)^3 (\textbf{r}_2^2 {{-}} 9 (\textbf{r}_2{{\cdot}} \textbf{R}_0)^2) {{+}} 
  21 (\textbf{r}_1{{\cdot}} \textbf{R}_0)^2 \textbf{r}_2{{\cdot}} \textbf{R}_0 (\textbf{r}_2^2 {{-}} 2 \textbf{r}_1{{\cdot}} \textbf{r}_2 \\ 
 & \, {{-}} 3 (\textbf{r}_2{{\cdot}} \textbf{R}_0)^2) {{+}} 
  \textbf{r}_2{{\cdot}} \textbf{R}_0 (6 \textbf{r}_1^2 \textbf{r}_1{{\cdot}} \textbf{r}_2 {{-}} 6 (\textbf{r}_1{{\cdot}} \textbf{r}_2)^2 {{+}} 
   \textbf{r}_1^2 ({{-}}3 \textbf{r}_2^2 {{+}} 7 (\textbf{r}_2{{\cdot}} \textbf{R}_0)^2)) \\ 
 & \, {{+}} 
  3 \textbf{r}_1{{\cdot}} \textbf{R}_0 (\textbf{r}_1^2 (\textbf{r}_2^2 {{-}} 7 (\textbf{r}_2{{\cdot}} \textbf{R}_0)^2) {{+}} 
   2 \textbf{r}_1{{\cdot}} \textbf{r}_2 ({{-}}\textbf{r}_2^2 {{+}} \textbf{r}_1{{\cdot}} \textbf{r}_2 {{+}} 7 (\textbf{r}_2{{\cdot}} \textbf{R}_0)^2)) \big) \\
 V_{5({\rm d \shortminus t})} &= \tfrac{3 q_A q_B}{8 R^7} \big( 5 \textbf{r}_1{{\cdot}} \textbf{r}_2 (\textbf{r}_1^4 {{+}} \textbf{r}_2^4 {{-}} 14 \textbf{r}_1^2 (\textbf{r}_1{{\cdot}} \textbf{R}_0)^2 {{+}} 21 (\textbf{r}_1{{\cdot}} \textbf{R}_0)^4 {{-}} 
   14 \textbf{r}_2^2 (\textbf{r}_2{{\cdot}} \textbf{R}_0)^2 \\ 
 & \, {{+}} 21 (\textbf{r}_2{{\cdot}} \textbf{R}_0)^4) {{-}} 
  7 \textbf{r}_1{{\cdot}} \textbf{R}_0 \textbf{r}_2{{\cdot}} \textbf{R}_0 (5 (\textbf{r}_1^4 {{+}} \textbf{r}_2^4) {{-}} 30 \textbf{r}_1^2 (\textbf{r}_1{{\cdot}} \textbf{R}_0)^2 {{+}} 
   33 (\textbf{r}_1{{\cdot}} \textbf{R}_0)^4 \\ 
 & \, {{-}} 30 \textbf{r}_2^2 (\textbf{r}_2{{\cdot}} \textbf{R}_0)^2 {{+}} 33 (\textbf{r}_2{{\cdot}} \textbf{R}_0)^4) \big) \\
 V_{5({\rm q \shortminus h})} &= \tfrac{15 q_A q_B}{16 R^7} \big( {{-}}168 (\textbf{r}_1{{\cdot}} \textbf{R}_0)^3 \textbf{r}_1{{\cdot}} \textbf{r}_2 \textbf{r}_2{{\cdot}} \textbf{R}_0 {{-}} 
  21 (\textbf{r}_1{{\cdot}} \textbf{R}_0)^4 (\textbf{r}_2^2 {{-}} 11 (\textbf{r}_2{{\cdot}} \textbf{R}_0)^2) \\ 
 & \, {{-}} 
  4 (\textbf{r}_1{{\cdot}} \textbf{r}_2)^2 (\textbf{r}_1^2 {{+}} \textbf{r}_2^2 {{-}} 7 (\textbf{r}_2{{\cdot}} \textbf{R}_0)^2) {{+}} 
  56 \textbf{r}_1{{\cdot}} \textbf{R}_0 \textbf{r}_1{{\cdot}} \textbf{r}_2 \textbf{r}_2{{\cdot}} \textbf{R}_0 (\textbf{r}_1^2 {{+}} \textbf{r}_2^2 \\ 
 & \, {{-}} 3 (\textbf{r}_2{{\cdot}} \textbf{R}_0)^2) {{-}} 
  \textbf{r}_1^2 (\textbf{r}_2^2 (\textbf{r}_1^2 {{+}} \textbf{r}_2^2) {{-}} 7 (\textbf{r}_1^2 {{+}} 2 \textbf{r}_2^2) (\textbf{r}_2{{\cdot}} \textbf{R}_0)^2 {{+}} 
   21 (\textbf{r}_2{{\cdot}} \textbf{R}_0)^4) \\ 
 & \, {{+}} 
  7 (\textbf{r}_1{{\cdot}} \textbf{R}_0)^2 (2 \textbf{r}_1^2 \textbf{r}_2^2 {{+}} \textbf{r}_2^4 {{+}} 4 (\textbf{r}_1{{\cdot}} \textbf{r}_2)^2 {{-}} 
   18 (\textbf{r}_1^2 {{+}} \textbf{r}_2^2) (\textbf{r}_2{{\cdot}} \textbf{R}_0)^2 {{+}} 33 (\textbf{r}_2{{\cdot}} \textbf{R}_0)^4) \big)  \\
 V_{5({\rm o \shortminus o})} &= \tfrac{5 q_A q_B}{4 R^7} \big( 2 (\textbf{r}_1{{\cdot}} \textbf{r}_2)^3 {{-}} 42 \textbf{r}_1{{\cdot}} \textbf{R}_0 (\textbf{r}_1{{\cdot}} \textbf{r}_2)^2 \textbf{r}_2{{\cdot}} \textbf{R}_0 {{+}} 
  3 \textbf{r}_1{{\cdot}} \textbf{r}_2 ({{-}}7 (\textbf{r}_1{{\cdot}} \textbf{R}_0)^2 (\textbf{r}_2^2 \\ 
 & \, {{-}} 9 (\textbf{r}_2{{\cdot}} \textbf{R}_0)^2) {{+}} 
   \textbf{r}_1^2 (\textbf{r}_2^2 {{-}} 7 (\textbf{r}_2{{\cdot}} \textbf{R}_0)^2)) {{-}} 
  21 \textbf{r}_1{{\cdot}} \textbf{R}_0 \textbf{r}_2{{\cdot}} \textbf{R}_0 (\textbf{r}_1^2 (\textbf{r}_2^2 {{-}} 3 (\textbf{r}_2{{\cdot}} \textbf{R}_0)^2) \\ 
 & \, {{+}} (\textbf{r}_1{{\cdot}} \textbf{R}_0)^2 ({{-}}3 \textbf{r}_2^2 {{+}} 
     11 (\textbf{r}_2{{\cdot}} \textbf{R}_0)^2)) \big) 
\end{align*} }

\section{Force balance for the quadrupole-quadrupole and octupole-octupole interactions}\label{appendix:B}

\begin{table}[h]
\centering
\caption{The proportionality coefficients $A_2^{\mu \omega}$, $A_{2 ({{ \rm q-q}) }}^{\mu \omega}$ and $A_3^{\mu \omega}$, $A_{3 ({{ \rm o-o}) }}^{\mu \omega}$ 
in comparison to the empirical reference values $A_l^{\rm ref} = R_{{ \rm vdW}}^{{ \rm ref}}/\alpha_{l,{ \rm ref}}^{2/7(l+1)}$. The used values of
$R_{{ \rm vdW}}^{{ \rm ref}}$, $\{ \mu , \omega \}$ and $\{ \alpha_{2,{ \rm ref}} , \alpha_{3,{ \rm ref}} \}$ are the same as in Table~\ref{tab:1}. The MARD is given in \%, whereas all the other quantities are in atomic units.}\label{tab:SI1}
\setlength{\tabcolsep}{6pt}
\begin{tabular}{c|ccc|ccc}
\hline\hline 
Atom & $A_2^{\mu \omega}$ & $A_{2 ({{ \rm q-q}) }}^{\mu \omega}$  & $A_2^{{\rm ref}}$ & $A_3^{\mu \omega}$ & $A_{3 ({{ \rm o-o}) }}^{\mu \omega}$  & $A_3^{{\rm ref}}$
\rule{0pt}{3.5ex} \\ [0.8ex]
\hline
He      & 2.10 & 1.95 & 2.43 & 1.93 & 1.82 & 2.24 \\
Ne      & 2.27 & 2.09 & 2.44 & 2.07 & 1.94 & 2.26 \\
Ar      & 2.18 & 2.08 & 2.44 & 2.06 & 2.01 & 2.27 \\
Kr      & 2.22 & 2.13 & 2.47 & 2.10 & 2.08 & 2.29 \\
Xe      & 2.20 & 2.14 & 2.45 & 2.10 & 2.13 & 2.27 \\
H       & 1.97 & 1.89 & 2.40 & 1.88 & 1.87 & 2.19 \\
Li      & 2.41 & 2.36 & 2.49 & 2.29 & 2.34 & 2.33 \\
Na      & 2.42 & 2.40 & 2.55 & 2.32 & 2.41 & 2.40 \\
K       & 2.47 & 2.49 & 2.54 & 2.38 & 2.54 & 2.41 \\
Rb      & 2.57 & 2.55 & 2.58 & 2.42 & 2.60 & 2.46 \\
Cs      & 2.56 & 2.62 & 2.58 & 2.46 & 2.69 & 2.48 \\
Ba      & 2.17 & 2.14 & 2.45 & 2.08 & 2.16 & 2.34 \\
Mg      & 2.21 & 2.22 & 2.42 & 2.13 & 2.27 & 2.32 \\
Ca      & 2.34 & 2.39 & 2.44 & 2.26 & 2.46 & 2.39 \\
Sr      & 2.41 & 2.47 & 2.49 & 2.32 & 2.54 & 2.43 \\
Ba      & 2.44 & 2.54 & 2.42 & 2.36 & 2.63 & 2.41 \\
\hline\hline
${\langle A_l \rangle}$ & \textbf{2.19} & \textbf{2.08} & \textbf{2.45} & \textbf{2.05} & \textbf{2.00} & \textbf{2.27} \\
${\sigma}$              & \textbf{0.16} & \textbf{0.22} & \textbf{0.06} & \textbf{0.17} & \textbf{0.40} & \textbf{0.08} \\
MARD & \textbf{7.13} & \textbf{11.49\,~} & \textbf{1.80} & \textbf{8.88} & \textbf{13.74\,~} & \textbf{3.90}
\end{tabular}
\end{table}

In order to demonstrate that the force balance is valid for each term in the multipole expansion, we calculate the proportionality coefficients $A_{2 ({ \rm q\shortminus q})} \equiv A_{2 ({ \rm q\shortminus q})}(\mu\omega, R_{\rm vdW})\, $ and $A_{3 ({ \rm o\shortminus o})} \equiv A_{3 ({ \rm o\shortminus o})}(\mu\omega, R_{\rm vdW})\,$ for the quadrupole-quadrupole and octupole-octupole interactions, which correspond to the ratios $R_{\rm vdW} / \alpha_{2}^{2/21}\,$ and $ R_{\rm vdW} / \alpha_{3}^{1/14}\,$, respectively. According to Eq.~\eqref{eq:alpha_QDO}, one has $\alpha_2 = (3/4)(\hbar/\mu\omega)\alpha_1$ and $\alpha_3 = (5/4)(\hbar/\mu\omega)^2\alpha_1$. By employing the aforementioned two relations and considering the quadrupole-quadrupole and octupole-octupole terms in Eqs.~(\ref{eq:exchange_forces}) and (\ref{eq:dispersion_forces}), we derive
\begin{equation}\label{eq:r/c(qq)}
A_{2 ({ \rm q\shortminus q})}^{\mu\omega} = \left(\tfrac{175}{1024} \right)^{\nicefrac{2}{21}} R_{\rm vdW}^{-\nicefrac{1}{21}} \left(\tfrac{\hbar}{\mu\omega}\right)^{\nicefrac{2}{7}} e^{\frac{4\mu \omega R_{\rm vdW}^2}{21\hbar}}
\end{equation} 
and
\begin{equation}\label{eq:r/c(oo)}
\begin{split}
A_{3 ({ \rm o\shortminus o})}^{\mu\omega} = \left(\tfrac{4851}{8192}\right)^{\nicefrac{1}{14}} R_{\rm vdW}^{-\nicefrac{1}{14}} \left(\tfrac{\hbar}{\mu\omega}\right)^{\nicefrac{2}{7}} e^{\frac{\mu \omega R_{\rm vdW}^2}{7\hbar}}\ .
\end{split}
\end{equation}
These two expressions can be compared to the coefficients $A_2^{\mu\omega}$ and $A_3^{\mu\omega}$, expressed by Eq.~\eqref{eq:r/c(dmult)}. The results are shown in Table~\ref{tab:SI1}, where we compare the results obtained from Eq.~\eqref{eq:r/c(dmult)} to those obtained from quadrupole-quadrupole and octupole-octupole interactions, expressed by Eqs.~\eqref{eq:r/c(qq)} and \eqref{eq:r/c(oo)}, respectively.
Remarkably, we found a good agreement between the proportionality coefficients $A_2^{\mu \omega}$ and $A_{2 ({ \rm q\shortminus q})}^{\mu \omega}$ with respect to the ratio of the vdW radius over the quadrupole polarizability; and between $A_3^{\mu \omega}$ and $A_{3 ({ \rm o\shortminus o})}^{\mu \omega}$ with respect to $R_{{ \rm vdW}}^{{ \rm ref}}/\alpha_{3,{ \rm ref}}^{1/14}$. For the noble gases, the mean values are $\langle A_{2 ({ \rm q\shortminus q}) }^{\mu \omega} \rangle = 2.08$~a.u. and $\langle A_{3 ({ \rm o\shortminus o})}^{\mu \omega} \rangle = 2.00$~a.u. with a deviation of $6.3 \% $ for the quadrupole-quadrupole interaction and $8.7 \% $ for the octupole-octupole interaction. 
In both cases, the mean values obtained form the high multipole terms (quadrupole-quadrupole and octupole-octupole) differ most from the ``universal'' values $A_2 = 2.45$~a.u. and $A_3 = 2.27$~a.u. with respect to the ones obtained from dipole-quadrupole and dipole-octupole interactions. Moreover, the error for $\langle A_{3 ({ \rm o\shortminus o}) }^{\mu \omega} \rangle$ is bigger than that for $\langle A_{2 ({ \rm q\shortminus q}) }^{\mu \omega} \rangle$. This means that, as expected, the lower-order contributions in the multipole expansion are more accurate with respect to the higher-order terms.

\end{document}